\documentclass[showpacs,preprintnumbers,amsmath,amssymb,nofootinbib,onecolumn,superscriptaddress,notitlepage]{revtex4-1}

\usepackage{amssymb}
\usepackage{graphicx}
\usepackage{amsmath}
\usepackage{multirow}
\usepackage{verbatim}
\usepackage{rotating}
\usepackage[usenames,dvipsnames]{color}
\usepackage[bookmarks=true, bookmarksnumbered=true, colorlinks=true,
  urlcolor=darkblue, citecolor=darkgreen, linkcolor=darkred,
  breaklinks=true]{hyperref} 
\usepackage{xspace}

\definecolor{darkblue}{rgb}{0,0,0.5}
\definecolor{darkgreen}{rgb}{0.1,0,0.3}
\definecolor{darkred}{rgb}{0.6,0,0}
\usepackage{tabularx}

\newcommand{\nc}{\newcommand}
\nc{\like}{\mathcal{L}}
\nc{\Emin}{{E_{\rm min}}}
\nc{\Emax}{{E_{\rm max}}}
\nc{\pdf}{\mathcal{P}}

\def\beq{\begin{equation}}
\def\eeq{\end{equation}}
\def\bea{\begin{eqnarray}}
\def\eea{\end{eqnarray}}

\begin{document}
\preprint{IFIC/16-64}

%%%%%%%%%%%%%%%%%%%%%%%%%%%%%%%%%%%%%%%%%%%%%

\title{Non-standard interactions with high-energy atmospheric neutrinos at  IceCube} 

\author{Jordi Salvado}
\email{jsalvado@ific.uv.es}
\affiliation{Instituto de F\'{\i}sica Corpuscular (IFIC)$,$
	CSIC-Universitat de Val\`encia$,$ \\  
	Apartado de Correos 22085$,$ E-46071 Valencia$,$ Spain}

\author{Olga Mena}
\email{omena@ific.uv.es}
\affiliation{Instituto de F\'{\i}sica Corpuscular (IFIC)$,$
 CSIC-Universitat de Val\`encia$,$ \\  Apartado de Correos 22085$,$ E-46071 Valencia$,$ Spain}

\author{Sergio Palomares-Ruiz}
\email{sergiopr@ific.uv.es}
\affiliation{Instituto de F\'{\i}sica Corpuscular (IFIC)$,$
 CSIC-Universitat de Val\`encia$,$ \\  
 Apartado de Correos 22085$,$ E-46071 Valencia$,$ Spain}
 
\author{Nuria Rius}
\email{nuria.rius@ific.uv.es}
\affiliation{Instituto de F\'{\i}sica Corpuscular (IFIC)$,$
 CSIC-Universitat de Val\`encia$,$ \\  
 Apartado de Correos 22085$,$ E-46071 Valencia$,$ Spain}

\begin{abstract}
\vspace{1cm}
Non-standard interactions in the propagation of neutrinos in matter can lead to significant deviations from expectations within the standard neutrino oscillation framework and atmospheric neutrino detectors have been considered to set constraints. However, most previous works have focused on relatively low-energy atmospheric neutrino data. Here, we consider the one-year high-energy through-going muon data in IceCube, which has been already used to search for light sterile neutrinos, to constrain new interactions in the $\mu\tau$-sector. In our analysis we include several systematic uncertainties on both, the atmospheric neutrino flux and on the detector properties, which are accounted for via nuisance parameters. After considering different primary cosmic-ray spectra and hadronic interaction models, we obtain the most stringent bound on the off-diagonal $\varepsilon_{\mu \tau}$ parameter to date, with the 90\% credible interval given by $-6.0 \times 10^{-3} < \varepsilon_{\mu \tau} < 5.4 \times 10^{-3}$. In addition, we also estimate the expected sensitivity after 10~years of collected data in IceCube and study the precision at which non-standard parameters could be determined for the case of $\varepsilon_{\mu \tau}$ near its current bound. 
\end{abstract}


\maketitle

\section{Introduction}
\label{sec:intro}

Neutrino oscillations have been robustly established over the past decades and this has been deservedly awarded during the last years. From neutrino oscillation experiments we know neutrinos have mass, which implies the first departure from the Standard Model (SM) of particle physics. Oscillation data provide information on the mixing angles and on the mass squared differences, which are, in the minimal and largely successful three-neutrino scenario, the solar mass splitting ($\Delta m_{12}^2\simeq 7.5\times 10^{-5}$~eV$^2$) and the atmospheric mass splitting ($|\Delta m_{23}^2|\simeq 2.5\times 10^{-3}$~eV$^2$)~\cite{Capozzi:2013csa, Forero:2014bxa, Gonzalez-Garcia:2014bfa}. Despite the enormous observational success achieved in constraining the leptonic mixing sector, there are still some unknowns in the neutrino mixing picture. Namely, the sign of the largest mass splitting remains unknown, as well as the octant of the mixing angle $\theta_{23}$ and the possible existence of leptonic CP violation. Neutrino physics has already entered into the high-precision measurements era and subleading effects due to exotic couplings, affecting neutrino production, propagation and/or detection processes, may also appear in the neutrino sector~\cite{Grossman:1995wx}. These, so-called, non-standard interactions (NSI) have been subject of extensive work in the past years (for recent reviews see, e.g., Refs.~\cite{Ohlsson:2012kf, Miranda:2015dra}), both from a pure theoretical perspective (see, e.g., Refs.~\cite{Honda:2007wv, Antusch:2008tz, Gavela:2008ra}) and with more phenomenological approaches, constraining their relative size with different experimental setups (see, e.g., Refs.~\cite{Berezhiani:2001rs, Berezhiani:2001rt, Guzzo:2001mi, Davidson:2003ha, Friedland:2004pp, Barranco:2005ps, Barranco:2007ej, Bolanos:2008km, Escrihuela:2009up, Biggio:2009kv, Biggio:2009nt, Forero:2011zz, Escrihuela:2011cf, Agarwalla:2012wf, Girardi:2014gna, Girardi:2014kca, Khan:2016uon}). Although constructing ($SU(2)_L \times U(1)_Y$ gauge-invariant) models with large neutrino NSI and consistent with all current experimental constraints, mainly from charged-lepton flavor-violating processes, requires a certain amount of fine-tuning~\cite{Gavela:2008ra}, they cannot be completely excluded. Therefore, from the phenomenological point of view, it is worth to exploit all available data to constrain neutrino NSI.

The relative size of NSI with respect to standard neutrino oscillations depends on the neutrino energy. At very low (sub-GeV) energies, the NSI terms are sub-dominant with respect to standard (vacuum) neutrino oscillations. At intermediate energies, $\mathcal{O}(1-10)$~GeV, NSI can interfere with the standard matter potential and vacuum oscillation terms, modifying the neutrino propagation through the Earth. At higher energies, NSI effects may dominate. Notice, however, that such an energy dependence is different if the NSI are due to light mediators (see, for instance, Refs.~\cite{Farzan:2015hkd, Farzan:2016wym}). In this case the effects depend on the high-energy, gauge-invariant, completion of each scenario. We will not consider this possibility, since our analysis is based on model-independent four-fermion effective operators, that we assume to be generated above the electroweak scale.

Therefore, exploiting the NSI energy dependence over a large range of energies and baselines seems a promising way of constraining these new potential neutrino interactions. As the presence of the NSI affects neutrino propagation in a medium, having a large range of available neutrino baselines crossing the Earth would help enormously in disentangling standard oscillations from NSI. Thus, atmospheric neutrinos provide a unique and ideal tool to test and constrain the size of NSI effects~\cite{Ma:1997ww, Brooijmans:1998py, GonzalezGarcia:1998hj, Lipari:1999vh, Fornengo:1999zp, Fornengo:2001pm, GonzalezGarcia:2004wg, Friedland:2004ah, Friedland:2005vy, GonzalezGarcia:2011my, Mitsuka:2011ty, Ohlsson:2013epa, Esmaili:2013fva, Chatterjee:2014gxa, Choubey:2014iia, Mocioiu:2014gua, Fukasawa:2015jaa, Thomas, Choubey:2015xha, Day:2016shw, DayN2016, Fukasawa:2016nwn}, as their spectrum covers a huge energy range ($\sim 0.1-10^5$~GeV) and, depending on their arrival direction, they may travel distances across the Earth ranging from tens to several thousands kilometers. Most works in the literature have focused on the capabilities of past and current~\cite{GonzalezGarcia:1998hj, Lipari:1999vh, Fornengo:1999zp, Fornengo:2001pm, GonzalezGarcia:2004wg, Friedland:2004ah, Friedland:2005vy, GonzalezGarcia:2011my, Mitsuka:2011ty, Esmaili:2013fva, Mocioiu:2014gua, Fukasawa:2015jaa, Thomas, Day:2016shw, DayN2016} or future detectors~\cite{Ohlsson:2013epa, Esmaili:2013fva, Chatterjee:2014gxa, Choubey:2014iia, Fukasawa:2015jaa, Choubey:2015xha, Fukasawa:2016nwn} using atmospheric neutrino events in the ${\mathcal O}(10~{\rm GeV})$ energy range, where interference effects may take place. In particular, the Super-Kamiokande (SK) collaboration (exploiting the sub-GeV, multi-GeV, stopping and through-going muon samples) obtained the most stringent bounds on the diagonal and off-diagonal NSI parameters in the $\mu\tau$-sector~\cite{Mitsuka:2011ty, Thomas} and recently, the IceCube Collaboration has also presented a preliminary analysis~\cite{DayN2016} using the DeepCore three-year muon disappearance result~\cite{Aartsen:2014yll}, with slightly more restrictive limits than SK.

With the development of neutrino observatories, NSI searches via atmospheric neutrino fluxes benefit from larger detector sizes (and consequently, larger atmospheric neutrino event samples) and major improvements in energy reconstruction at neutrino energies above ${\mathcal O}(10~{\rm GeV})$, and up to ${\mathcal O}(10~{\rm PeV})$, although they have higher energy thresholds. This has been the main goal of Refs.~\cite{Ohlsson:2013epa, Esmaili:2013fva, Choubey:2014iia, Mocioiu:2014gua, Day:2016shw, DayN2016}, where the ice \v{C}erenkov IceCube neutrino observatory and/or its low energy extensions, the DeepCore or the future PINGU detectors, have been considered as the ideal targets where to test neutrino NSI, exploiting atmospheric neutrino fluxes. On the other hand, although at high-energies the standard neutrino oscillation phase, which is inversely proportional to the neutrino energy, is very small, in the presence of NSI, oscillations are not suppressed with energy and they only depend on the baseline. The data from the 79-string configuration in IceCube~\cite{Aartsen:2013jza} was used to set constraints on NSI in Ref.~\cite{Esmaili:2013fva}. 

Here we perform an analysis of the NSI effects on the propagation of high-energy atmospheric neutrinos by considering the publicly available IceCube one-year upgoing muon sample~\cite{TheIceCube:2016oqi}, referred to as IC86 (IceCube 86-string configuration), which contains 20145 muons detected over a live time of 343.7 days. We focus on the high-energy region of the atmospheric neutrino spectrum, and thus our results are complementary to those of previous analyses of IceCube data~\cite{Esmaili:2013fva, Day:2016shw, DayN2016}, some of them dealing exclusively with the low-energy atmospheric neutrino sample observed at the DeepCore detector~\cite{Day:2016shw, DayN2016}. In order to perform the analysis, we use the public IceCube Monte Carlo\footnote{\url{https://icecube.wisc.edu/science/data/IC86-sterile-neutrino}} that models the detector realistically and allows us to relate physical quantities, as the neutrino energy and direction, to observables, as the reconstructed muon energy and zenith angle. To account for some possible systematic uncertainties on the atmospheric neutrino flux, neutrino parameters and detector properties, we also include a number of nuisance parameters. We obtain the most stringent limits to date on the off-diagonal NSI parameter $\varepsilon_{\mu \tau}$. Finally, we also present a forecast of the sensitivity to NSI from future high-energy atmospheric neutrino data. We simulate 10~years of collected data in IceCube and assess how the bounds would improve and how well the presence of NSI could be determined, in case they exist.

The paper is organized as follows.  In Sec.~\ref{sec:forma} we briefly review the NSI formalism relevant for the data we consider, i.e., for high-energy atmospheric neutrinos crossing the Earth, and describe the main features of the NSI effects. Then, in Sec.~\ref{sec:data} we describe the data we use and explicitly show the potential effects of NSI on this type of observations. In Sec.~\ref{sec:results}, we first describe the likelihood and the different systematic uncertainties included in the analysis, presenting then the current bounds on NSI using the one-year through-going muon IceCube data. We finish that section by discussing the prospects for future limits with improved statistics (10 years of data) and we summarize our findings in Sec.~\ref{sec:conclusions}.

\section{Formalism}
\label{sec:forma}

We consider neutrino NSI that are generated by new physics above the electroweak scale, so that at low center-of-mass energies, $E \ll m_W$ (or, equivalently, $E \ll m_X$, where $m_X$ is the mass of the heavy mediator), they can be described via model-independent four-fermion effective operators. These can be of neutral current (NC) type~\cite{Grossman:1995wx}, 
\beq
\label{eq:NC}
{\cal L}^{\rm NC}_{\rm NSI} = - 2 \sqrt{2} \, G_F \, \varepsilon^{fP}_{\alpha \beta} \, 
 (\bar{\nu}_\alpha \gamma_{\rho} L \nu_\beta) (\bar{f} \gamma^{\rho}P  f) ~,
\eeq
where $\varepsilon^{fP}_{\alpha \beta}$ are the NC NSI parameters (by hermiticity $\varepsilon^{fP}_{\alpha \beta} = (\varepsilon^{fP}_{\beta \alpha})^*$), $P=\{L,R\}$ (with $L$ and $R$ the left and right quirality projectors) and $f$ is any SM fermion, as well as of charged-current (CC) type~\cite{Grossman:1995wx, GonzalezGarcia:2001mp}, 
\bea
\label{eq:NC}
{\cal L}^{{\rm CC}\ell}_{\rm NSI} & = & - \ 2 \sqrt{2} \, G_F \, \varepsilon^{\delta \sigma P}_{\alpha \beta} \, (\bar \nu_\alpha \gamma_{\rho} L \nu_\beta) (\bar \ell_\delta \gamma^{\rho}P  \ell_\sigma)  ~, \\
{\cal L}^{{\rm CC}q}_{\rm NSI} & = & - \ 2 \sqrt{2} \, G_F \, \varepsilon^{qq'P}_{\alpha \beta} \, 
(\bar \nu_\alpha \gamma_{\rho} L \ell_\beta) (\bar{q} \gamma^{\rho}P  q') + h.c. ~, 
\eea
where $\varepsilon^{\delta \sigma P}_{\alpha \beta}$ and $\varepsilon^{qq'P}_{\alpha \beta}$ are the leptonic and hadronic CC NSI parameters (for the leptonic case, $\delta = \sigma$ corresponds to NC NSI), $\ell_\beta$ is a charged lepton of flavor $\beta$, $q$ is a down-type quark and $q'$ an up-type quark. In what follows, we neglect possible CP violation in the new interactions (this has been considered in different contexts~\cite{GonzalezGarcia:2001mp, Gago:2001xg, Gago:2001xg, Ota:2001pw, Ota:2002na, Bekman:2002zk, Campanelli:2002cc, Hattori:2002uw, Friedland:2004pp, GonzalezGarcia:2004wg, Blennow:2007pu, Kopp:2007mi, FernandezMartinez:2007ms, Kopp:2007ne, Ribeiro:2007ud, Liao:2008qu, Kopp:2008ds, Blennow:2008ym, Winter:2008eg, Altarelli:2008yr, Kikuchi:2008vq, Ohlsson:2008gx, Gago:2009ij,GonzalezGarcia:2011my, Coloma:2011rq, Blennow:2015nxa, deGouvea:2015ndi, Coloma:2015kiu, Forero:2016cmb, Masud:2016bvp, Blennow:2016etl, Ge:2016dlx}), so we take all NSI parameters $\varepsilon^{fP}_{\alpha \beta}$, $\varepsilon^{\delta \sigma P}_{\alpha \beta}$ and $\varepsilon^{qq'P}_{\alpha \beta}$ to be real.

In the literature, NC NSI are frequently called matter NSI, since they modify neutrino propagation through matter, while the CC ones are referred to as production and detection NSI.  Moreover, given that the neutrino flavor is always tagged through the flavor of the charged lepton associated with it, in the presence of CC NSI the neutrino flavor basis is not well-defined~\cite{Grossman:1995wx, GonzalezGarcia:2001mp}, since the neutrino detected or produced in association with a charged lepton does not necessarily share its flavor. In this case, flavor conversion is present at the interaction level, and the standard oscillation formulae become more cumbersome~\cite{GonzalezGarcia:2001mp, Ota:2001pw}. Model-independent bounds on both, NC and CC NSI have been derived in Ref.~\cite{Biggio:2009nt}, where it was found that, in general, the limits on production and detection NSI are one order of magnitude more stringent than those on matter NSI. Model-dependent bounds in several new physics scenarios~\cite{Grossman:1995wx, GonzalezGarcia:2001mp} also indicate that constraints on CC NSI are typically much more stringent. Therefore, we shall neglect CC NSI and concentrate only on NC NSI in the following.

The standard evolution Hamiltonian for neutrinos includes the coherent forward scattering on fermions of the type $f$, $\nu_\alpha + f \to \nu_\beta + f$, given by the matter interaction potential (defined in Eq.~(\ref{eq:hnsi}) below), which affects neutrino oscillations. However, neutrinos propagating through the Earth can also interact inelastically with matter, either via CC or NC processes. As the neutrino-nucleon cross section increases with energy, for energies above $\sim$TeV, the neutrino flux gets attenuated~\cite{Gandhi:1995tf, Gandhi:1998ri}. Whereas in the case of $\nu_e$'s and $\nu_\mu$'s, the neutrino flux is absorbed via CC interactions and redistributed (degraded in energy) via NC interactions~\cite{Berezinsky:1986ij}, in the case of $\nu_\tau$'s, there is another effect. Unlike what happens for $\nu_e$ and $\nu_\mu$ CC interactions, where charged leptons are quickly brought to rest and do not contribute to the high-energy neutrino flux, the tau leptons produced after $\nu_\tau$ CC interactions can decay before being stopped, so $\nu_\tau$'s are not absorbed, but the flux gets regenerated (at lower energies)~\cite{Halzen:1998be, Iyer:1999wu, Dutta:2000jv, Hettlage:2001yf, Hussain:2003vi, Yoshida:2003js, Reya:2005vh}.  Thus, for each $\nu_\tau$ which is absorbed via CC interactions, another $\nu_\tau$ with lower energy is produced, and the Earth does not become opaque to high-energy $\nu_\tau$'s. In addition, secondary $\nu_e$'s and $\nu_\mu$'s are also produced after tau leptons decay into leptonic channels~\cite{Beacom:2001xn, Dutta:2002zc}. For high-energy neutrinos, oscillation, attenuation and regeneration effects occur simultaneously when they travel across the Earth, and the evolution equations should, in principle, include them. Notice that conventional neutrino oscillation analyses do not take into account attenuation and regeneration effects, which is a good approximation, provided the energy of the detected neutrinos is low enough. Nevertheless, this is not the case for the high-energy IceCube sample of atmospheric neutrinos we consider in this work, for which attenuation needs to be included. On the contrary, for atmospheric neutrinos, the effects of $\nu_\tau$ regeneration and production of secondary $\nu_e$ and $\nu_\mu$ fluxes are very small. The explanation is two fold. On one hand, $\nu_\tau$'s are very rarely produced after cosmic-ray interactions in the atmosphere, and therefore the atmospheric $\nu_\tau$ flux is negligible. On the other hand, these effects are only relevant for very hard spectra. Therefore, for the sake of computational time, we shall not include $\nu_\tau$ regeneration in this study, which nevertheless implies negligible corrections.

In what follows, we use the density matrix, $\rho(E_\nu, x) = \nu(E_\nu, x) \otimes \nu(E_\nu, x)^\dagger$, formalism, where $E_\nu$ is the neutrino energy and $x$ the path variable. In the case of neglecting neutrino regeneration, the density matrix for neutrinos traversing the Earth obeys the evolution equation~\cite{GonzalezGarcia:2005xw}
 \beq
 \label{eq:rho}
 \frac{d \rho(E_\nu, x)}{dx} = - i [H(E_\nu,x ),\rho(E_\nu, x)] - \sum_{\alpha} \frac{1}{2 \, \lambda_\alpha(E_\nu ,x)} \{\Pi_\alpha(E_\nu), \rho(E_\nu, x)\} + \int_{E_\nu}^{\infty} \rho(E_\nu', x) \frac{1}{n_N(x)} \, \frac{d\sigma_{\rm NC} (E_\nu', E_\nu)}{dE_\nu} dE_\nu' ~,
 \eeq
where $\Pi_\alpha$ is the $\nu_\alpha$ projector, $\lambda_\alpha(E_\nu, x) = 1/[n_N(x) \, \sigma^{\rm tot}_\alpha (E_\nu)]$ is the attenuation length of $\nu_\alpha$, with $n_N(x)$ the nucleon number density in the Earth\footnote{In general, one has to include all possible targets, as electrons, but at the energies of interest in this work, interactions with electrons have a negligible effect in the attenuation of the neutrino flux.} and $\sigma^{\rm tot}_\alpha (E_\nu)$ the $\nu_\alpha$ total (CC+NC) cross section, and $d\sigma_{\rm NC}/dE_\nu$ is the differential NC cross section. The first term on the right-hand side represents neutrino oscillations, the second term neutrino absorption and the third term the redistribution of the flux due to NC interactions.

In the presence of NSI, the effective Hamiltonian that controls neutrino propagation in matter can be written as
\beq
\label{eq:hnsi}
H(E_\nu,x) = \frac{1}{2 E_\nu} U M^2 U^\dagger + {\rm diag}(V_e,0,0)  + \sum_f V_f  \, \varepsilon^{fV}  \ , 
\eeq
where $U$ is the PMNS mixing matrix, $M^2 = {\rm diag}(0, \Delta m_{21}^2, \Delta m_{31}^2)$,  with  $\Delta m_{ij}^2 \equiv m_i^2 - m_j^2$ the neutrino mass square differences and $V_e (x)= \sqrt{2} \, G_F \,  n_e(x)$ corresponds to the standard neutrino flavor potential in matter, with $n_e(x)$ the electron number density.  The effect of NSI is encoded in the last term of Eq.~(\ref{eq:hnsi}), where $V_f (x)=  \sqrt{2} \, G_F\,  n_f(x)$, with $n_f(x)$ the number density of fermion $f$, and $\varepsilon^{fV}$ is the matrix in lepton flavor space that contains the vector combination of the NSI chiral parameters, $\varepsilon^{fV}_{\alpha \beta}  =  \varepsilon^{fR}_{\alpha \beta}  +  \varepsilon^{fL}_{\alpha \beta}$. As in the case of SM interactions, the matter term for antineutrinos changes sign and one has to make the substitution $V_f \to - V_f$ (and $U \to U^*$). On the other hand, it is convenient to define effective NSI parameters for a given medium (from now on we omit for simplicity the $x$ dependence of the number densities) by normalizing the fermion number density, $n_f$, to the density of $d$-quarks, $n_d$,
\beq 
\varepsilon_{\alpha \beta}  \equiv \sum_{f}  \frac{n_f}{n_d} \, \varepsilon^{fV}_{\alpha \beta} ~, 
\eeq
so that $\sum_f V_f  \varepsilon^{fV} \equiv  V_e \, r \, \varepsilon = V_d \, \varepsilon$, and $r = n_d/n_e$. For the Earth, $n_n \approx n_p$ and therefore, $r \approx 3$.

Given the current constraints on the electron neutrino NSI parameters $\varepsilon_{e \alpha}$, and, for energies above the resonance in the $13$-sector ($E_\nu \gtrsim 20$~GeV), one of the mass eigenstates (mostly $\nu_e$ or $\bar \nu_e$) decouples from the other two states. Therefore, the $\nu_e \to \nu_\mu$ transition does not affect the IceCube events as it is strongly suppressed and moreover, the initial atmospheric $\nu_e$ and $\bar \nu_e$ fluxes are much smaller than the $\nu_\mu$ and $\bar \nu_\mu$ fluxes. Thus, we can approximately describe the evolution of the system as that of a two-neutrino system, focusing on the $23$-block of the evolution Hamiltonian, Eq.~(\ref{eq:hnsi}). Recall that neutrino oscillations are only sensitive to the difference in the diagonal effective parameters, i.e.,  $\varepsilon' = \varepsilon_{\tau \tau} - \varepsilon_{\mu \mu}$, which modifies the oscillation probability due a change of the effective matter density felt by neutrinos, while the off-diagonal term, $\varepsilon_{\mu \tau}$, shifts the effective mixing angle in the medium. The diagonal parameter $\varepsilon'$ characterizes the lack of universality of NC in the $\mu\tau$-sector, and the off-diagonal $\varepsilon_{\mu \tau}$ quantifies the strength of flavor changes in NC interactions.

Before discussing the main features of the transition probabilities at high energies, we would like to point out that the effects of NSI in high-energy atmospheric neutrinos in IceCube differ from the standard approach at lower energies in two ways: 
\begin{enumerate}
\item[1)]
Usually, only NSI of neutrinos with quarks and leptons of the first generation can be bounded, via the 
$V_f  \, \varepsilon^{fV}$ contributions to the matter Hamiltonian and via the $\varepsilon^{u d V}$ contributions to CC interactions with pions and nucleons, in addition to the $\varepsilon^{e \mu V}$ contributions at production via muon decay.  However, very energetic neutrinos ($E_\nu \gtrsim$~TeV) can {\it see} the strange quark contribution inside nucleons, since for such high energies the strange quark parton distribution function is not negligible. As a consequence, there is an effective energy dependence of production and detection NSI terms (if NSI do not affect all quark flavors with the same strength) through the different contribution of the corresponding parton distribution at different energies. As mentioned above, here we do not consider CC NSI and, as done in the literature, we assume the NC NSI parameters to be equal for all quarks inside the nucleons. Relaxing these assumptions, IceCube data could also be used to bound strange quark CC NSI with neutrinos, by properly taking into account the energy dependence of the $s$ quark contribution to the parton distribution functions.

\item [2)]
Matter NSI could also modify the total inelastic scattering cross section, by altering the NC cross section, and thus, the absorption term in Eq.~(\ref{eq:rho}). Attenuation is negligible at low energies, but it is relevant for the high-energy IceCube neutrinos, so there could be some sensitivity to the  presence of NSI. However, CC interactions are $\sim$~2.4 times larger than the NC cross sections~\cite{Gandhi:1995tf, Gandhi:1998ri}, so the latter dominate the absorption term and the effect of NC NSI can be safely neglected. Moreover, CC NSI could also be present, but for the values of the CC NSI parameters currently allowed, the NSI effects on attenuation would be very small, implying corrections to the results presented in this work below the percent level. On the other hand, by modifying the NC cross section, NSI would also alter the degradation in energy of the neutrino flux while crossing the Earth. Given the fact that for atmospheric neutrinos this effect is subdominant, we also neglect the NSI correction on the last term of Eq.~(\ref{eq:rho}).
\end{enumerate}

In our calculations, we solve numerically the full three-neutrino evolution equation, using the values of the neutrino mixing parameters from Ref.~\cite{Gonzalez-Garcia:2014bfa} assuming normal hierarchy and including the effects mentioned above. To compute the neutrino propagation through the Earth, we use the publicly available libraries SQuIDS and $\nu$-SQuIDS~\cite{squids,nusquids} in the Trunk version found in the repositories~\cite{squidsweb, nusquidsweb}. Nevertheless, in order to understand the effects of the diagonal ($\varepsilon'$) and off-diagonal ($\varepsilon_{\mu \tau}$) NSI parameters in the energy range we consider, it is interesting to note that, for atmospheric neutrinos, the interplay of neutrino oscillations and attenuation in the Earth can be well described by an overall exponential suppression in the oscillated fluxes, i.e.,
\beq
\label{eq:oscatt}
\phi_\alpha (E_\nu, \theta_z) = \phi_\mu^0 (E_\nu, \theta_z) \, P\left(\nu_\mu \to \nu_\alpha; E_\nu, L(\theta_z)\right) \, \exp \{- \int_{0}^{L(\theta_z)} dx/\lambda_\alpha (E_\nu, x)\} ~, 
\eeq
where $\phi_\mu^0$ is the atmospheric $\nu_\mu$ (or $\bar \nu_\mu$) flux before entering the Earth and $L(\theta_z)$ is the baseline across the Earth in a direction with zenith angle $\theta_z$. In this way, it is illustrative to study analytically the oscillation probabilities in the approximation of constant matter density (assuming constant NSI parameters), i.e., the solution of the evolution equation neglecting attenuation and energy degradation, regeneration and secondary production (see Ref.~\cite{Esmaili:2013fva} for a detailed discussion) and for constant density. In this case, the two-neutrino oscillation probability after propagating over a distance $L$, $P(\nu_\mu \to \nu_\tau) = 1 - {\rm Tr}\{\Pi_\mu \rho\}$, is given by~\cite{Coleman:1998ti}
\beq
\label{eq:P2nu}
P(\nu_\mu \to \nu_\tau) = \sin^2 2\theta_{\rm mat} \, \sin^2\left(\frac{\Delta m_{31}^2 \, L}{4 \, E_\nu} \, R\right) ~,
\eeq
where 
\bea
\label{eq:R}
R^2 & = & 1 + R_0^2 + 2 \, R_0 \, \cos 2 (\theta_{23} - \xi) ~, \\
\sin^2 2\theta_{\rm mat} & = & \frac{\left(\sin 2\theta_{23} + R_0 \, \sin 2 \xi\right)^2}{R^2} ~, 
\eea
with
\bea
\label{eq:R0}
R_0 & = & \frac{\phi_{\rm mat}}{\phi_{\rm vac}} = \frac{V_{\rm NSI} \, L/2}{\Delta m^2_{31} \, L /4  E_\nu}~,  \\
V_{\rm NSI} & = & V_d \, \sqrt{4 \, \varepsilon_{\mu \tau}^2 + \varepsilon'^{2}} ~, \\
\sin 2 \xi & = & \frac{2 \, \varepsilon_{\mu \tau}}{\sqrt{4 \, \varepsilon_{\mu \tau}^2 + \varepsilon'^{2}}} ~. 
\eea

\begin{figure}
	\centering
	\includegraphics[width=0.49\textwidth]{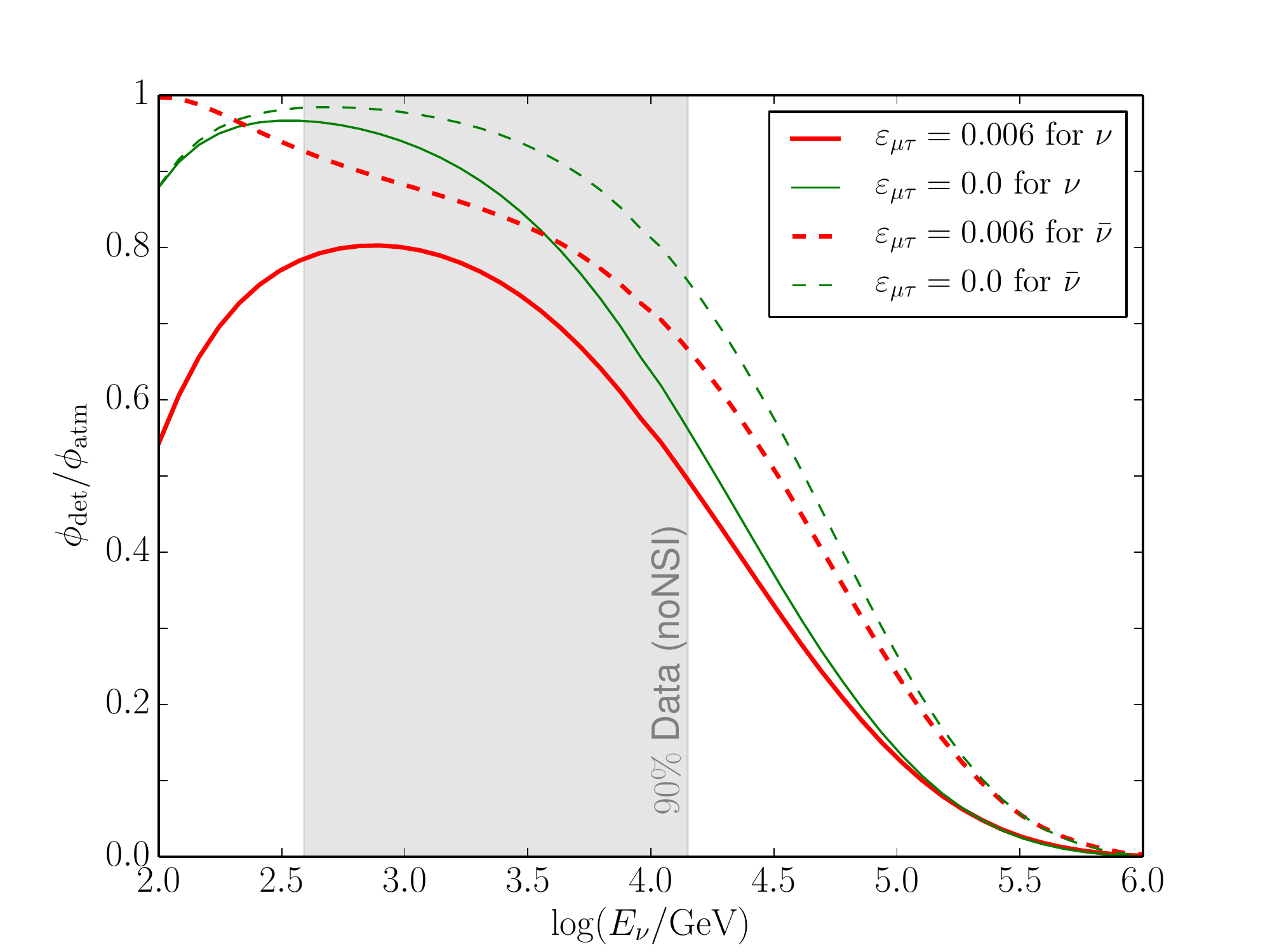}  
	\includegraphics[width=0.49\textwidth]{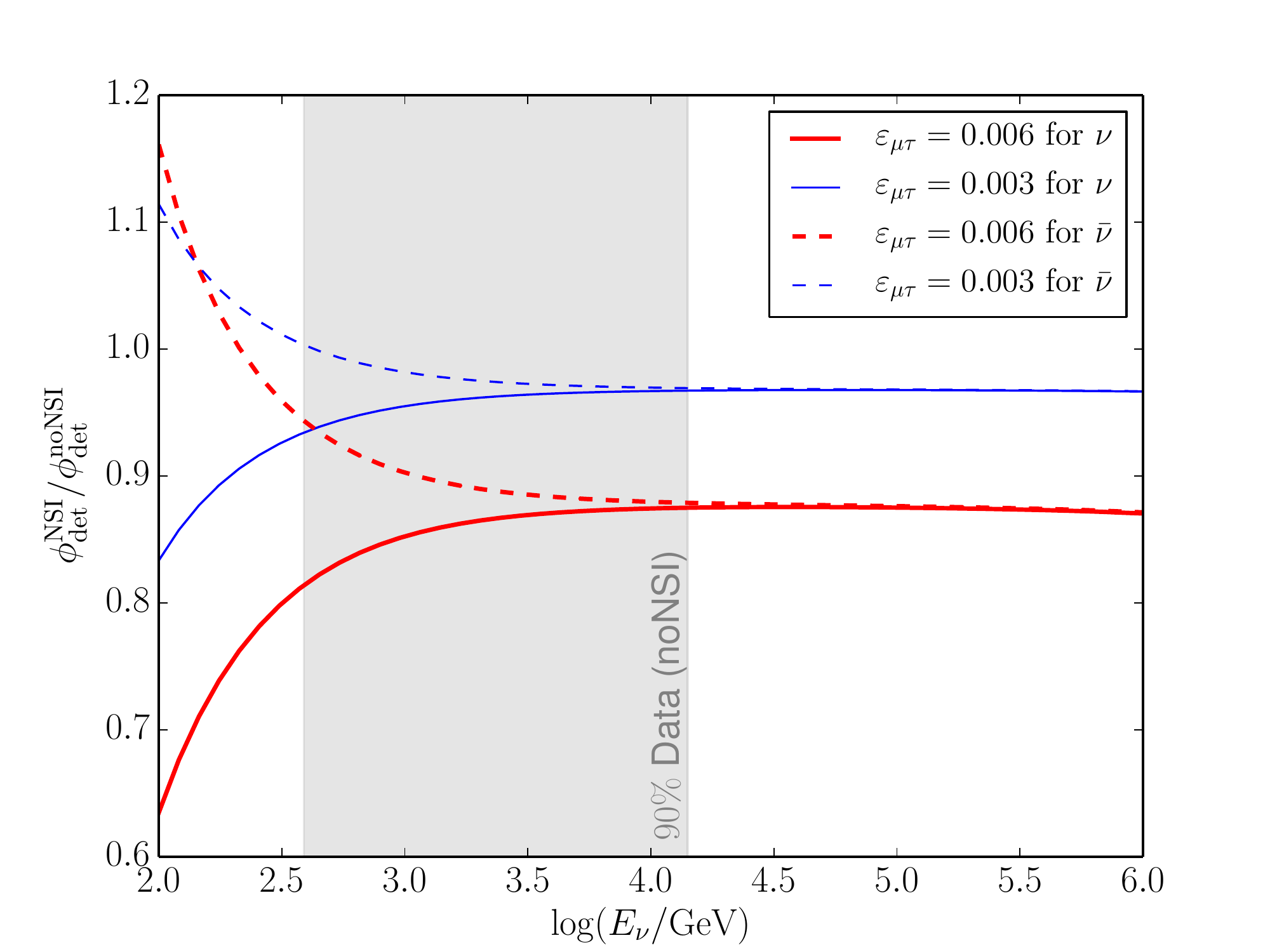}  
	\caption{{\it Left panel:} Comparison of the ratios of propagated to unpropagated atmospheric $\nu_\mu$ (solid lines) and $\bar \nu_\mu$ (dashed lines) fluxes for values of the NSI off-diagonal parameter $\varepsilon_{\mu \tau} = 0.006$ (thick red lines) and $\varepsilon_{\mu\tau} = 0$ (thin green lines). {\it Right panel:} Comparison of the ratios of atmospheric $\nu_\mu$ (solid lines) and $\bar \nu_\mu$ (dashed lines) fluxes at the detector (after propagation) with NSI to those without NSI, for $\varepsilon_{\mu\tau} = 0.003$ (thin blue lines) and 0.006 (thick red lines). In both panels, the ratios are shown for $\cos \theta_z = -1$ and we have chosen $\varepsilon' = 0$. For illustration, we show the gray area, which corresponds to the energy interval that produced 90\% of the events in the entire sample considered here in the absence of NSI effects.}
	\label{fig:fratio_1D}
\end{figure}

For the energies we consider in this work ($E_\nu > 100$~GeV), neutrino oscillations in vacuum are suppressed for baselines comparable to or smaller than the Earth diameter, $\phi_{\rm vac} \equiv \Delta m_{31}^2 L / 4 E_\nu \ll 1$. Therefore, in the case in which the vacuum and matter terms in the oscillation phase are of the same order of magnitude, i.e., $R_0 = {\cal O}(1)$, the transition probability approximately reads
\beq
\label{eq:vacmat}
P(\nu_\mu \to \nu_\tau) \simeq \left(\sin 2\theta_{23} + R_0 \, \sin 2 \xi\right)^2 \, \phi_{\rm vac}^2 = \left(\sin 2\theta_{23} \, \frac{\Delta m^2_{31}}{2 \, E_\nu} +  2 \, V_d \, \varepsilon_{\mu \tau}\right)^2 \left(\frac{L}{2}\right)^2  ~.
\eeq
Considering normal hierarchy, for neutrinos ($R_0 \, \sin 2\xi > 0$) this probability is enhanced and thus, the $\nu_\mu$ flux is suppressed with respect to the case without NSI (vacuum oscillations, $R_0 \, \sin 2 \xi = 0$), whereas for antineutrinos ($R_0 \, \sin 2\xi < 0$) it is the other way around. It is interesting to note that in \v{C}erenkov detectors like IceCube, neutrinos cannot be distinguished from antineutrinos, so this effect tends to partially cancel out, as we will see below. These differences can be clearly seen for neutrino energies $E_\nu = {\cal O}(100~{\rm GeV})$ in both panels of Fig.~\ref{fig:fratio_1D}, where we show the effect of neutrino propagation through the Earth with and without NSI. This regime corresponds to the low-energy part of the event sample we consider, as illustrated by the gray area in both panels, which represents the energy interval that produced 90\% of the events in the entire sample assuming no NSI (5\% upper cut and 5\% lower cut). Notice that in this section we use true neutrino energies and zenith angles, whereas the IceCube events are described by reconstructed variables which differ from the true ones, mainly in the case of the reconstructed energy, which is always smaller than the true neutrino energy. 

In the left panel of Fig.~\ref{fig:fratio_1D}, we plot the ratio of the propagated to unpropagated atmospheric $\nu_\mu$ (solid lines) and $\bar \nu_\mu$ (dashed lines) fluxes with (thick red lines) and without (thin green lines) NSI, whereas in the right panel, we show the ratios of atmospheric $\nu_\mu$ and $\bar \nu_\mu$ fluxes at the detector (after propagation) with NSI to those without NSI. In the left panel we show the case of $\varepsilon_{\mu \tau} = 0.006$ (thick red lines) and no NSI, $\varepsilon_{\mu \tau} = 0$, (thin green lines) and in the right panel we depict the ratios for two representative values of the NSI off-diagonal parameter $\varepsilon_{\mu \tau} = 0.003$ (thin blue lines) and 0.006 (thick red lines). In both panels, we consider muon neutrinos and antineutrinos traversing the entire Earth, $\cos \theta_z = -1$, and we have set $\varepsilon' = 0$ ($\sin^2 2 \xi = 1$). In the left panel, in addition to the effect of oscillations at low energies (even without NSI), we can clearly see the effect of attenuation (and the subdominant degradation in energy via NC interactions) at higher energies. These curves approximately represent the product of the oscillation and attenuation terms in the right-hand side of Eq.~(\ref{eq:oscatt}). The differences between the neutrino and antineutrino results are two fold: at low energies the oscillation probabilities and, at all energies in the plot, the total cross sections, and thus, the attenuation factors, are different for neutrinos and antineutrinos. On the other hand, the effect of attenuation is factored out in the right panel, which approximately represents the ratio of the survival probabilities with and without NSI. Note that, at first order, the transition probability for energies $E_\nu = {\cal O}(100~{\rm GeV})$, Eq.~(\ref{eq:vacmat}), is independent of $\varepsilon'$ and thus, there is little sensitivity to the diagonal NSI parameter. In the limit of $\varepsilon' \gg \varepsilon_{\mu \tau}$ ($\sin 2 \xi \simeq 0$), at high energies, vacuum mimicking is realized~\cite{Yasuda:2001va}, but the oscillation phase is suppressed. Hence, there is significantly more sensitivity to $\varepsilon'$ for $E_\nu < 100$~GeV~\cite{GonzalezGarcia:1998hj, Lipari:1999vh, Fornengo:1999zp, Fornengo:2001pm, GonzalezGarcia:2004wg, Friedland:2004ah, Friedland:2005vy, Escrihuela:2011cf, GonzalezGarcia:2011my, Ohlsson:2013epa, Esmaili:2013fva, Chatterjee:2014gxa, Choubey:2014iia, Mocioiu:2014gua, Fukasawa:2015jaa, Choubey:2015xha, Fukasawa:2016nwn}.

\begin{figure}
	\centering
	\includegraphics[width=0.49\textwidth]{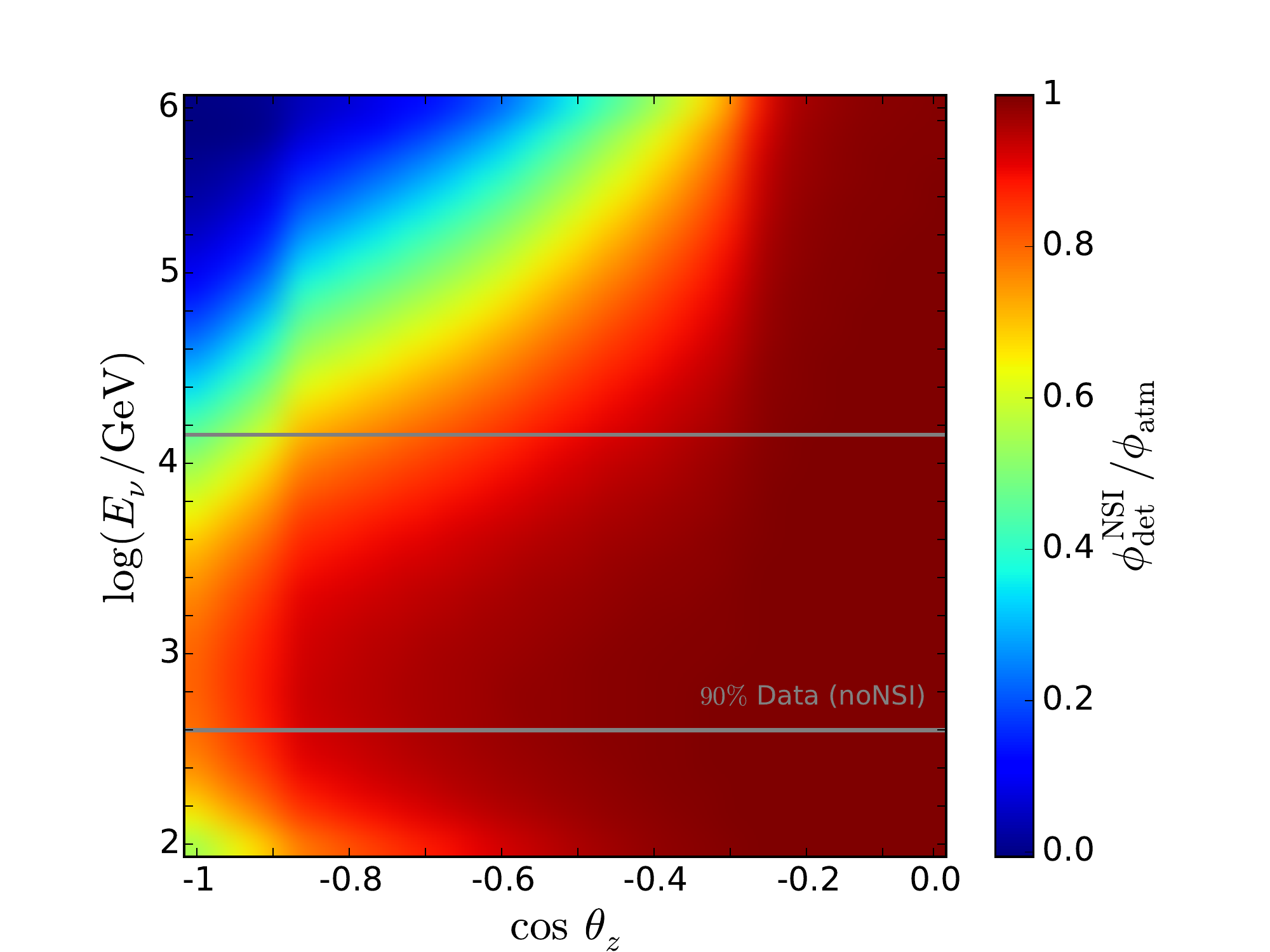} \includegraphics[width=0.49\textwidth]{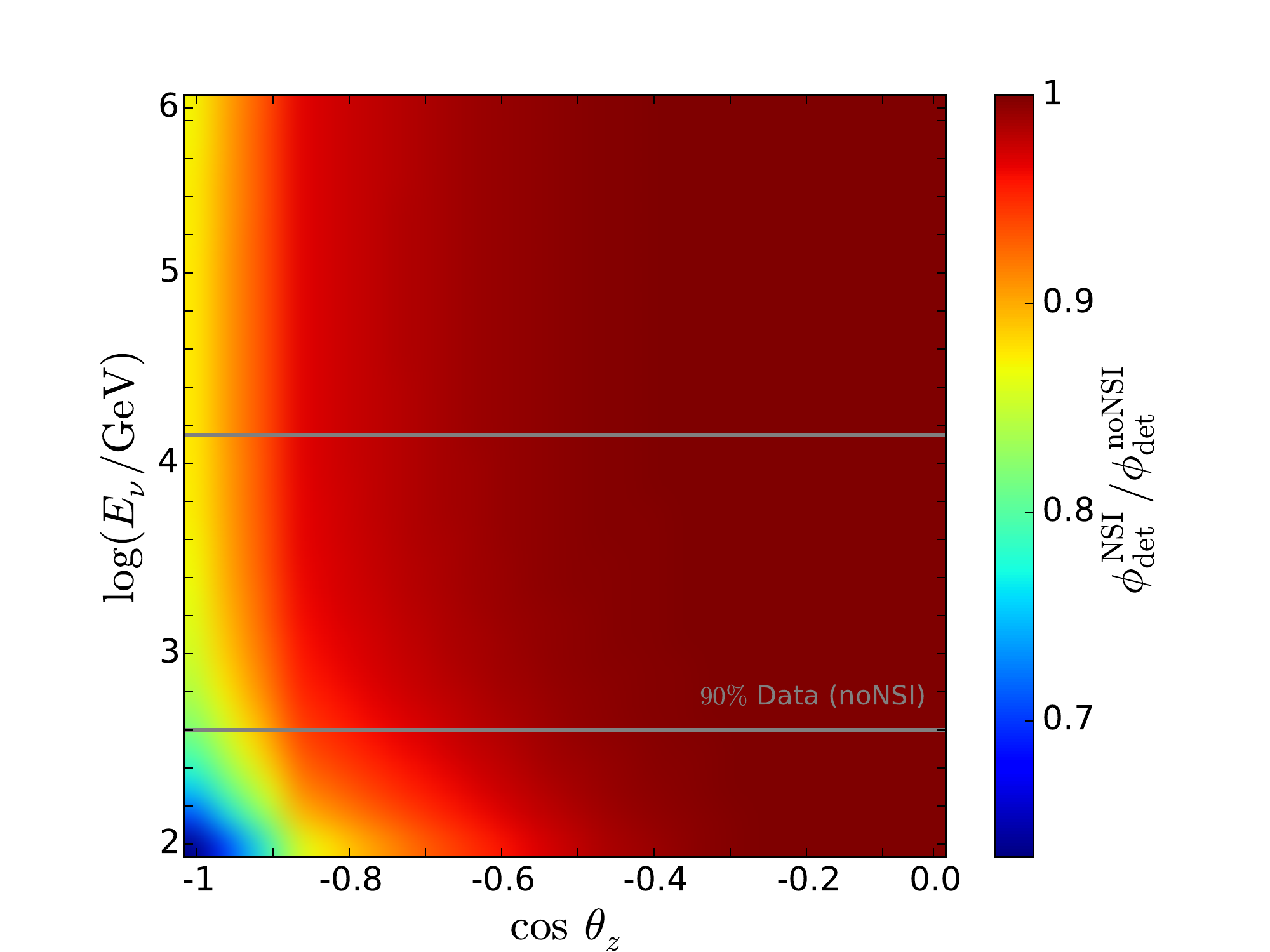}  
	\caption{{\it Left panel:} Ratio of propagated to unpropagated atmospheric $\nu_\mu$ fluxes as a function of the neutrino energy and the zenith angle. {\it Right panel:} Ratio of atmospheric $\nu_\mu$ fluxes at the detector (after propagation) with NSI to those without NSI. In both panels we choose the NSI off-diagonal parameter to be $\varepsilon_{\mu \tau} = 0.006$ and the diagonal parameter to be $\varepsilon' = 0$. We also show two gray lines, which bound the energy interval in which 90\% of the events in the entire sample ( assuming no NSI) are produced.} 
	\label{fig:fratio_2D}
\end{figure}

On the other hand, in the high-energy limit, the matter term dominates over vacuum oscillations, i.e., $ \phi_{\rm mat} \gg \phi_{\rm vac}$ or $R_0 \gg 1$. In this regime, the two-neutrino oscillation probability, Eq.~(\ref{eq:P2nu}), is approximately given by 
\beq
\label{eq:analyticform} 
P(\nu_\mu \to \nu_\tau) \simeq \sin^2 2 \xi \, \sin^2 \phi_{\rm mat}  \ , 
\eeq
where 
\beq 
\phi_{\rm mat}  = \frac{V_d L}{2} \sqrt{4 \, \varepsilon_{\mu \tau}^2 + \varepsilon'^2} \simeq
30 \left(\frac{\rho}{8 \, {\rm g/cm^3}} \right) 
\left(\frac{L}{2 \, R_{\oplus}} \right) \sqrt{4 \, \varepsilon_{\mu \tau}^2 + \varepsilon'^2}  ~, 
\eeq
with $R_{\oplus}$ the radius of the Earth. Then, for $\phi_{\rm mat} \ll1$, the transition probability   
\beq 
P(\nu_\mu \to \nu_\tau) \simeq \left(\sin^2 2\xi\right) \, \phi_{\rm mat}^2 =  (\varepsilon_{\mu \tau} \, V_d \, L)^2 
\eeq
is proportional to $\varepsilon_{\mu \tau}^2$ and becomes independent of $\varepsilon'$~\cite{Gago:1999ws}, and the same result holds for antineutrinos. This is also clearly seen in the high-energy regime shown in the right panel of Fig.~\ref{fig:fratio_1D}, where one can see that the neutrino and antineutrino ratios of (approximately) oscillation probabilities coincide. As a consequence, the high-energy IceCube atmospheric neutrino sample cannot significantly constrain the diagonal NSI parameter $\varepsilon'$, so in our analysis we use information on $\varepsilon'$ based on the SK limits~\cite{Mitsuka:2011ty} (see below), obtained from its potential effects at lower energies using the zenith distribution.

Finally, in Fig.~\ref{fig:fratio_2D}, in addition to the energy dependence, we also show the dependence on the zenith angle of the ratios depicted in Fig.~\ref{fig:fratio_1D}. The effect of NSI in the neutrino propagation through the Earth is illustrated for $\varepsilon_{\mu\tau}=0.006$ and $\varepsilon'=0$. In the left panel, analogously to the left panel of Fig.~\ref{fig:fratio_1D}, we show the ratio between the initial atmospheric $\nu_\mu$ flux and the $\nu_\mu$ flux in the detector. In the right panel, we isolate the effect produced by the NSI on the oscillation probabilities, as in the right panel of Fig.~\ref{fig:fratio_1D}, displaying the ratio between the final fluxes for $\varepsilon_{\mu\tau}=0.006$ and $\varepsilon_{\mu\tau}=0.$, i.e., with and without NSI. Note that the left vertical axis ($\cos \theta_z = -1$) in both panels corresponds to the red solid lines in both panels of Fig.~\ref{fig:fratio_1D}.  We clearly see the well-known effects of attenuation that shift to higher energies for more horizontal trajectories (left panel) and the NSI-induced oscillations of the atmospheric $\nu_\mu$ flux, which represent a flux suppression that, at high energies, only depends on the zenith angle (right panel).

\section{Data description}
\label{sec:data}

\begin{figure}
	\centering
	\includegraphics[width=0.49\textwidth]{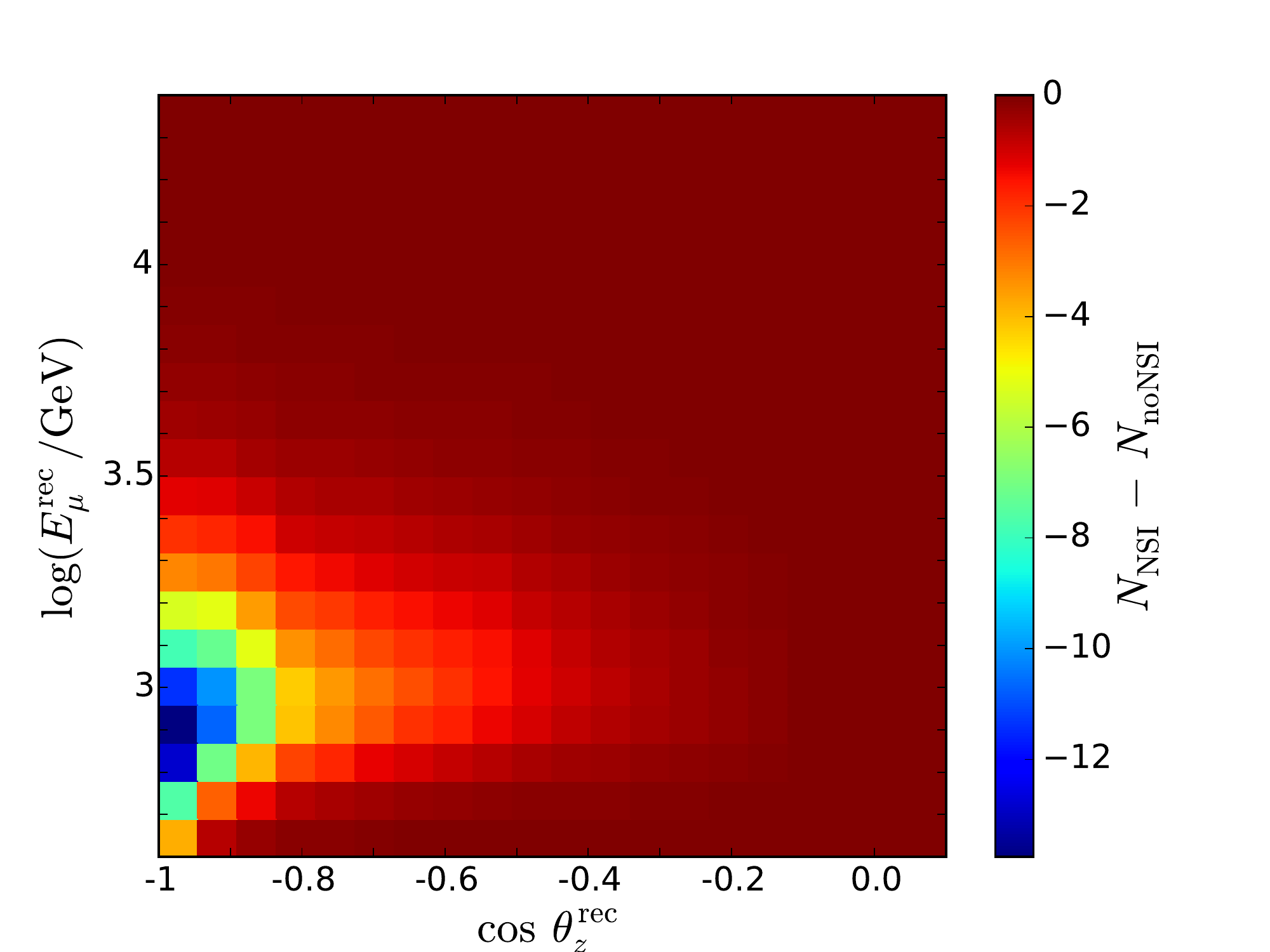} \includegraphics[width=0.49\textwidth]{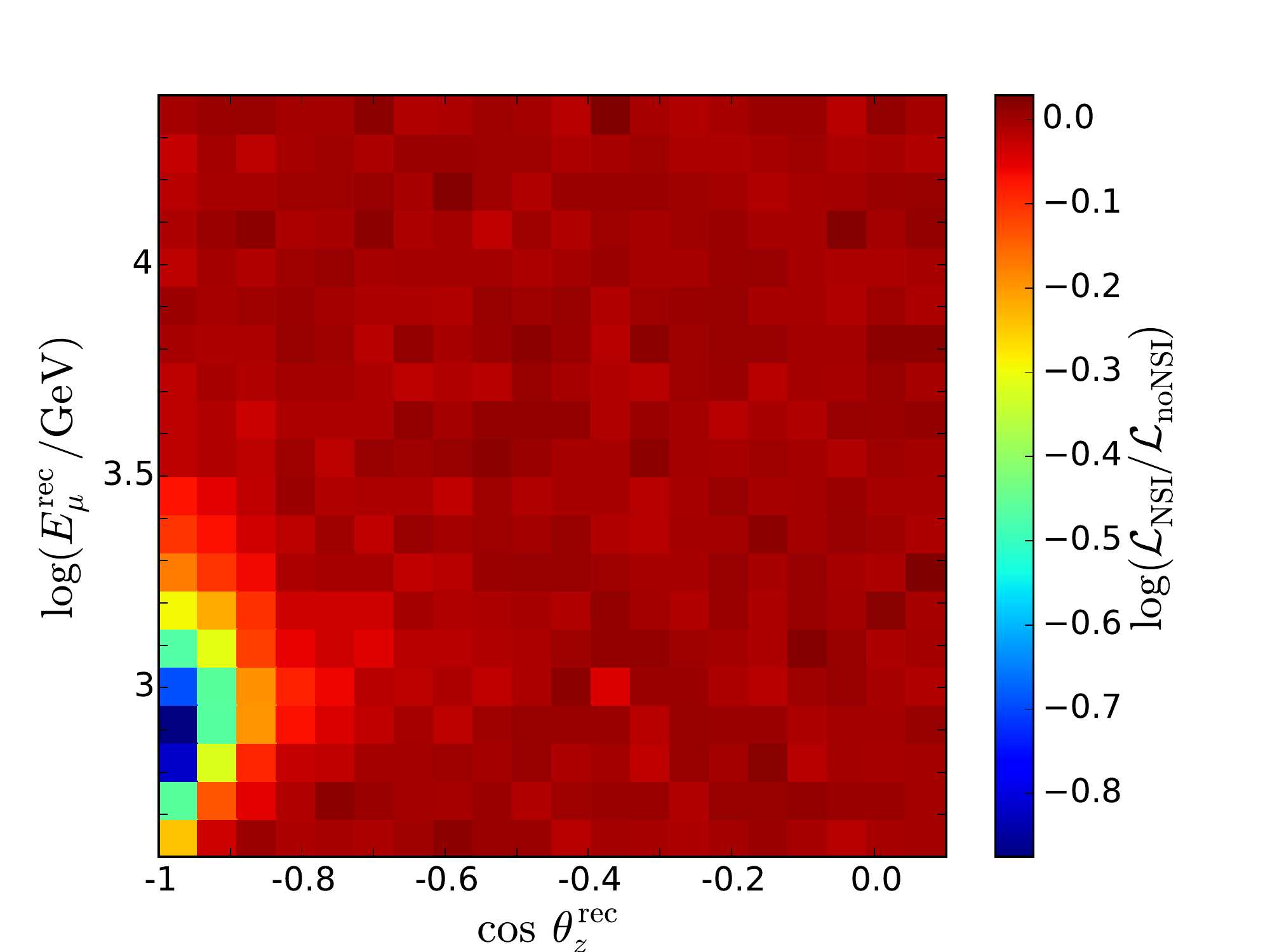}  
	\caption{{\it Left panel:} Difference of the expected number of events (from neutrinos and antineutrinos) between the case of $\varepsilon_{\mu\tau} = 0.006$ and $\varepsilon_{\mu\tau}=0$, as a function of reconstructed energy and zenith angle. {\it Right panel:} Statistical pulls as a function of reconstructed energy and zenith angle for the same set of parameters. In both panels we set $\varepsilon' = 0$.}
	\label{fig:DeltaEv}
\end{figure}

The IceCube data we consider in this paper is the same sample used to search for light sterile neutrino signatures~\cite{TheIceCube:2016oqi}. It contains 20145 events detected during 343.7 days of live data in the period 2011-2012 using the full IceCube 86-string configuration. These events correspond to upgoing neutrinos from the Northern hemisphere, which are dominantly produced by atmospheric $\nu_\mu$ and $\bar \nu_\mu$ CC interactions with nucleons of the material surrounding the detector, the so-called through-going muon tracks. The contamination from other sources is found to be below the 0.1\% level~\cite{TheIceCube:2016oqi}. The reconstructed muon energies in the detector of this sample lie in the range $E^{\rm rec}_\mu \simeq (300~{\rm GeV} - 20~{\rm TeV})$ and the neutrino energies mostly contributing to these events are indicated by the gray bands (lines) in Fig.~\ref{fig:fratio_1D} (Fig.~\ref{fig:fratio_2D}).

Through-going track events are produced after $\nu_\mu$ and $\bar \nu_\mu$ CC interactions produce muons outside the instrumented volume, that traverse the detector while depositing energy along their trajectory (the track). At these energies, muons travel along (almost) the same direction of the parent neutrino, which is reconstructed with very good angular resolution (within one degree or better, i.e., $\sigma_{\cos \theta_z} \simeq 0.005 - 0.015$~\cite{TheIceCube:2016oqi}). On the other hand, due to radiative losses, the fact that the position of the interaction vertex is unknown implies a large uncertainty in the estimation of the initial muon energy, which in turn is always smaller than the incoming neutrino energy. The muon energy when entering the detector is estimated based on the energy losses along the track~\cite{Aartsen:2013vja} with a resolution of $\sigma_{\log(E_\mu/{\rm GeV})} \sim 0.5$~\cite{TheIceCube:2016oqi}.

For our analysis, we use the high-statistics Monte Carlo released by the IceCube collaboration along with the data, which allows us to relate the true variables (neutrino energy and direction) to the reconstructed observables (deposited energy and track zenith angle) and to do a realistic treatment of the detector systematic uncertainties, which are described below. 

In order to understand the different features in the neutrino propagation induced by the NSI effects, and discussed in detail in previous sections, we simulate 1000 realizations of mock data corresponding to one year of observation. In the left panel of Fig.~\ref{fig:DeltaEv}, we show the expected difference in the number of events between the hypotheses without NSI ($\varepsilon_{\mu\tau} = \varepsilon' = 0$) and that in which NSI are included, with $\varepsilon_{\mu\tau} = 0.006$ (and $\varepsilon' = 0$), as a function of the reconstructed muon energy $E^{\rm rec}_\mu$ and zenith angle $\theta^{\rm rec}_z$. It is clear from that panel that the largest differences in the expected number of events with and without NSI occur for neutrinos crossing the core of the Earth with energies $\sim$TeV. 

Although Fig.~\ref{fig:DeltaEv} clearly illustrates the region in the parameter space which is sensitive to the NSI effects, we also quantify it statistically by defining a Poisson likelihood for each $i$-th bin (defined by an interval in $E^{\rm rec}_\mu$ and $\theta^{\rm rec}_z$)
\beq
\label{eq:Lsim}
\mathcal{L}_i = \frac{e^{-\langle N^{\rm sim}_i\rangle} {\langle N^{\rm sim}_i\rangle}^{N^{\rm sim}_i}}{N^{\rm sim}_i!} ~,
\eeq
where $\langle N^{\rm sim}_i \rangle$ ($N^{\rm sim}_i$) refers to the average over all realizations of the number of simulated events (number of simulated events for an individual realization) in the $i$-th bin. On the right panel of Fig.~\ref{fig:DeltaEv}, we show the expected average over all realizations of the log-likelihood difference between the null (assuming no NSI) and the NSI hypotheses (for $\varepsilon_{\mu\tau} = 0.006$ and $\varepsilon' = 0$). Notice that the two panels of Fig.~\ref{fig:DeltaEv} look very similar, which indicates that the impact of NSI is what pulls the statistical significance, rather than the higher statistics around the horizon, with shows a negligible dependence on NSI effects. Therefore, as already anticipated, the most sensitive region in the reconstructed variables is that corresponding to neutrinos that travel through the core of the Earth, i.e., $\cos \theta^{\rm rec}_z \lesssim -0.8$, with reconstructed energies for which the data sample has the higher statistics, i.e., $E^{\rm rec}_\mu \sim \cal{O} ({\rm TeV})$. This is expected, as the NSI effects turn out to be approximately energy independent.

\begin{figure}
	\centering
	\includegraphics[width=0.32\textwidth]{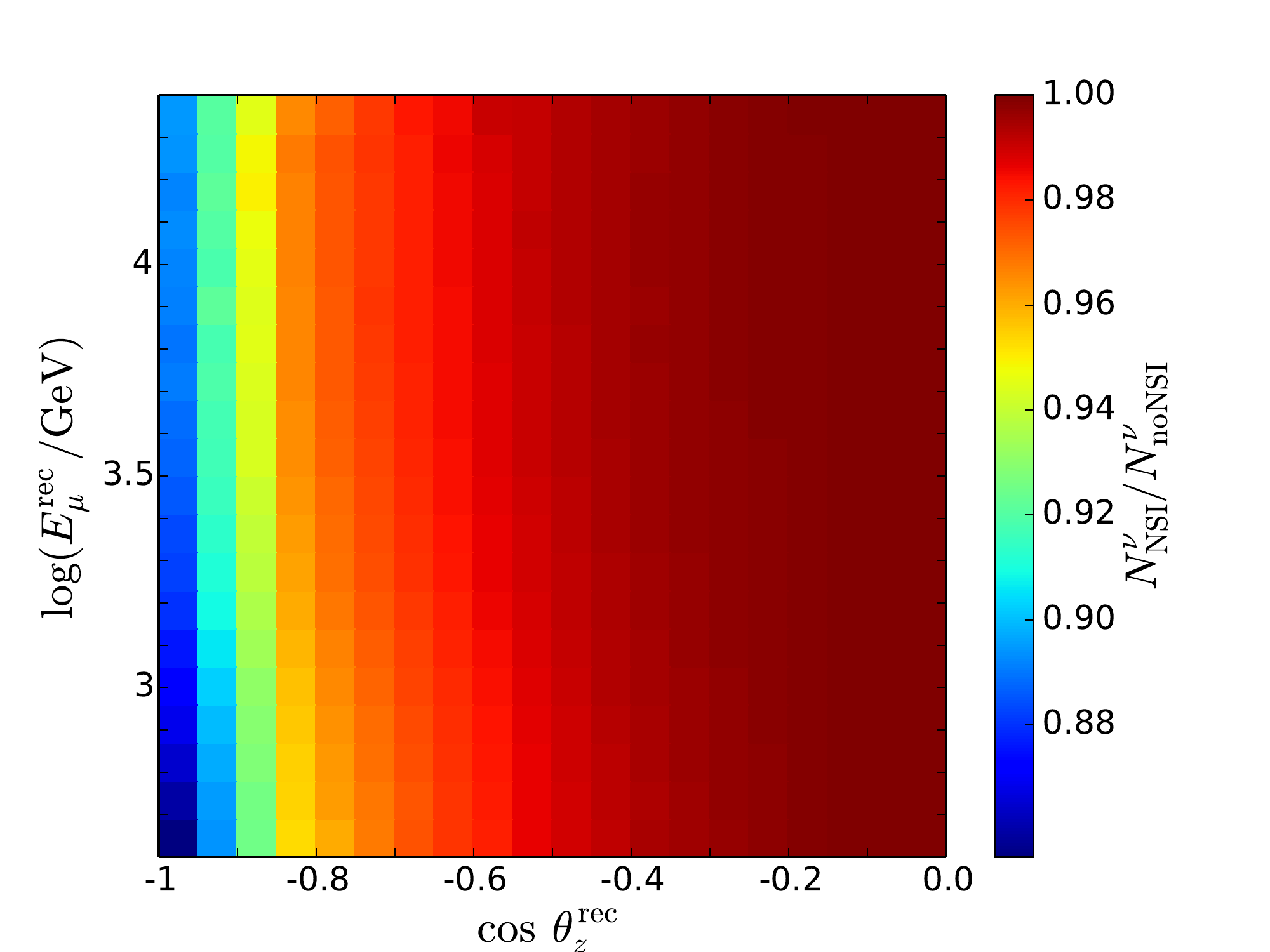} \includegraphics[width=0.32\textwidth]{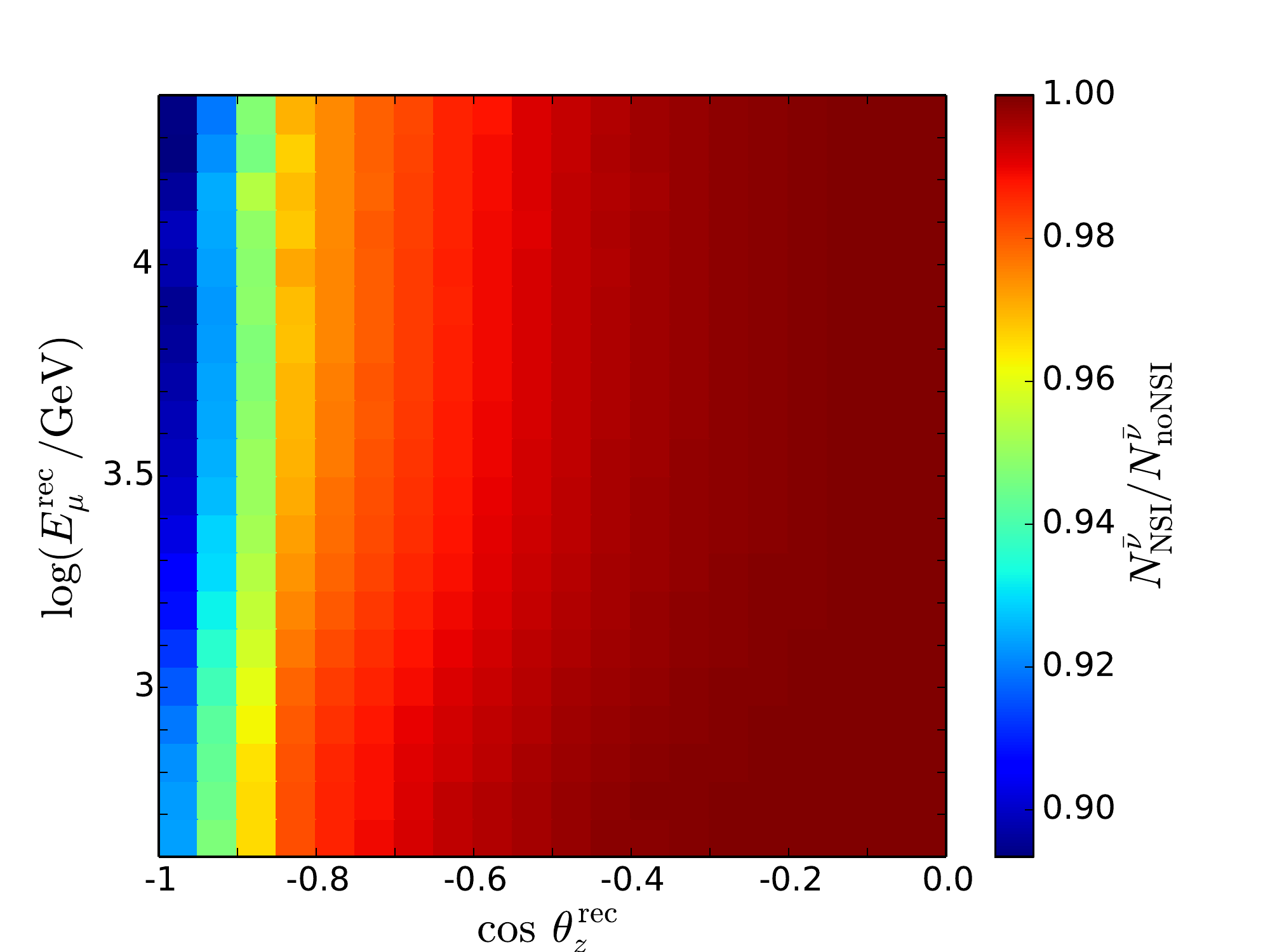}
	\includegraphics[width=0.32\textwidth]{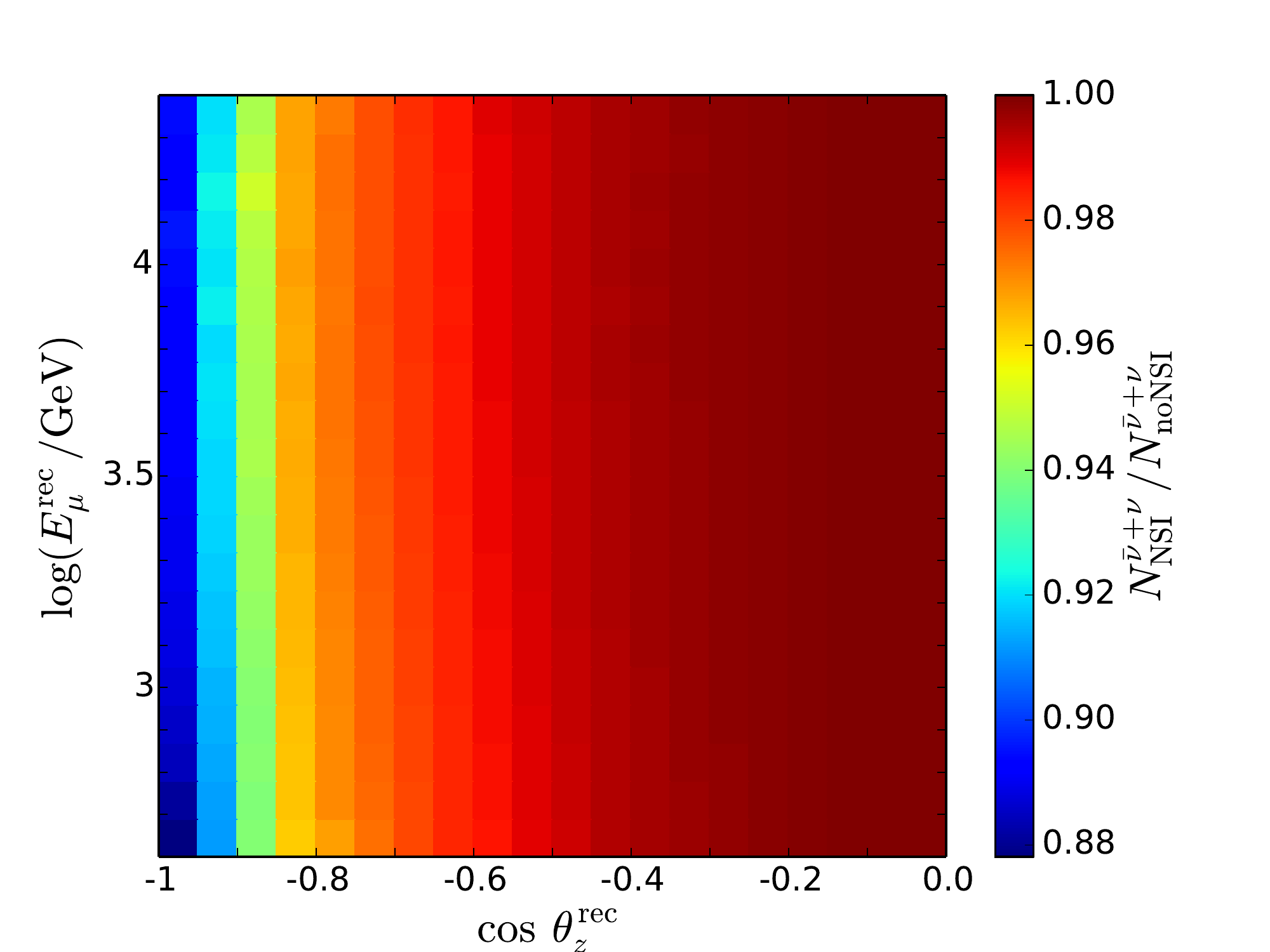}  
	\caption{{\it Left panel:} Ratio of the number of events produced by neutrinos including NSI ($\varepsilon_{\mu \tau} = 0.006$ and $\varepsilon' = 0$) to that without NSI ($\varepsilon_{\mu \tau} = 0$ and $\varepsilon' = 0$), as a function of the reconstructed muon energy and zenith angle. {\it Middle panel:} Same but for antineutrinos. {\it Right panel:} Same but for the total number of events, i.e., neutrino plus antineutrino events.}
	\label{fig:DeltaEvfrac}
\end{figure}

Indeed, this energy independence can be clearly noticed from the results shown in the right panel of Fig.~\ref{fig:DeltaEvfrac}, where we depict the ratio of neutrino plus antineutrino events including NSI ($\varepsilon_{\mu \tau} = 0.006$ and $\varepsilon' = 0$) to the number of events without NSI (i.e., with $\varepsilon_{\mu \tau} = 0$ and $\varepsilon' = 0$). In analogy to the right panel of Fig.~\ref{fig:fratio_1D}, in Fig.~\ref{fig:DeltaEvfrac}, we also show the ratios of neutrino (left panel) and antineutrino (middle panel) events including NSI ($\varepsilon_{\mu \tau} = 0.006$ and $\varepsilon' = 0$) to either neutrino or antineutrino events without NSI ($\varepsilon_{\mu \tau} = 0$ and $\varepsilon' = 0$). As expected, when NSI are at play, the ratio of events grows with energy for neutrinos and decreases with energy for antineutrinos in a very similar manner, and, consequently, once these two contributions are summed up (representing the measurable quantity in the IceCube neutrino telescope), their energy dependence approximately cancels out.

\section{Results}
\label{sec:results}

In this section we describe the results arising from our analyses. Firstly, we describe the different ingredients that enter into the definition of the likelihood and then we show the results obtained with the current one-year through-going muon IceCube data~\cite{TheIceCube:2016oqi}. Finally, we also perform forecast analyses with 10~years of simulated data considering two different hypotheses, with or without NSI. \\

\begin{table}[t]
\setlength{\tabcolsep}{0.5em}
{\renewcommand{\arraystretch}{1.2}		
	\begin{tabular}{ c || c | c | c | l}
		Parameter & Default value & Range & Prior & Description\\
		\hline
		$\varepsilon_{\mu\tau}$ & {\bf 0.006} & $[-1,1]$ & Flat & NSI flavor off-diagonal term\\
		$\varepsilon'$ & {\bf 0} & $[-1,1]$ & Gaussian: $\sigma=0.04$ & NSI flavor diagonal term\\
		$N$ & {\bf 1} & $[0.5,2.0]$ & Flat & Normalization of the energy spectrum\\
		$\pi/K$ & {\bf 1} & $[0.7,1.5]$ & Gaussian: $\sigma=0.10$  & Pion-to-kaon ratio contribution\\
		$\Delta\gamma$ & {\bf 0} & $[-0.2,0.2]$ & Gaussian: $\sigma=0.05$ & Tilt of the energy
		spectrum\\
		${\rm DOM}_{\rm eff}$ & {\bf 0.99} & $[0.90,1.19]$ & Flat & Optical efficiency\\
		$\Delta m^2_{31}/10^{-3}$ [eV$^2$] & {\bf 2.484} & $[2.3, 2.7]$ & Gaussian: $\sigma=0.048$ & Atmospheric mass square difference\\
		$\theta_{23}$ [$^\circ$] & {\bf49.3} & $[43.0, 54.4]$ & Gaussian: $\sigma=1.7$ & Atmospheric mixing angle\\
	\end{tabular}}
	\caption{Parameters, default values for plots (using the HG-GH-H3a primary cosmic-ray flux and the QGSJET-II-4 hadronic model), their range of variation and priors (flat or Gaussian) for the different systematics considered in our statistical analysis.}
	\label{tab:params}
\end{table}

\begin{table}[t]
\setlength{\tabcolsep}{0.5em}
{\renewcommand{\arraystretch}{1.2}	
  \begin{tabular}{ c || c c || c  c || c  c || c  c }
     & \multicolumn{2}{c||}{HG-GH-H3a + QGSJET-II-4} & \multicolumn{2}{c||}{HG-GH-H3a + SIBYLL2.3} & \multicolumn{2}{c||}{ZS + QGSJET-II-4} & \multicolumn{2}{c}{ZS + SIBYLL2.3} \\ \hline
     Parameter & \hspace{4mm} Mean & Std. dev. & \hspace{3mm} Mean & Std. dev. & \hspace{1mm} Mean & Std. dev. & \hspace{1mm} Mean & Std. dev. \\
    \hline
        $\varepsilon_{\mu\tau}$  &  \hspace{4mm} -0.0004 & 0.0034 & \hspace{3mm} 0.0001 & 0.0035 & \hspace{1mm} -0.0005 & 0.0036 & \hspace{1mm} -0.0002 & 0.0035 \\
        $\varepsilon'$                 &  \hspace{4mm} 0.000  & 0.047  & \hspace{3mm} -0.003 & 0.045 & \hspace{1mm} 0.002 & 0.046 & \hspace{1mm} 0.001 & 0.046 \\
        $N$                                &  \hspace{4mm} 1.013  & 0.056  & \hspace{3mm} 0.911  & 0.051 & \hspace{1mm} 1.257 & 0.066 & \hspace{1mm} 1.123  & 0.063 \\
        $\pi/K$                           &  \hspace{4mm} 1.078  & 0.084  & \hspace{3mm} 1.059  & 0.080 & \hspace{1mm} 1.073 & 0.080 & \hspace{1mm} 1.067 & 0.083 \\
        $\Delta\gamma$             &  \hspace{4mm} -0.050 & 0.013 & \hspace{3mm} -0.092 & 0.013 & \hspace{1mm} 0.066 & 0.012 & \hspace{1mm} 0.102 & 0.012 \\
        ${\rm DOM}_{\rm eff}$   &  \hspace{4mm} 0.9869 & 0.0064 & \hspace{3mm} 0.9863 & 0.0061 & \hspace{1mm} 0.9910 & 0.0061& \hspace{1mm} 0.9885 & 0.0058 \\
        $\Delta m^2_{31}/10^{-3}$ [eV$^2$]   &  \hspace{4mm} 2.484 & 0.046 & \hspace{3mm} 2.485 & 0.047 & \hspace{1mm} 2.487 & 0.044 & \hspace{1mm} 2.480 & 0.043 \\
        $\theta_{23}$ [$^\circ$] & \hspace{4mm} 49.3 & 1.8 & \hspace{3mm} 49.3 & 1.7 & \hspace{1mm} 49.3 & 1.7 & \hspace{1mm} 49.2 & 1.7
  \end{tabular}}
  \caption{Mean value and standard deviation for the parameters and systematics of this analysis, for each of the four combinations of primary cosmic-ray flux and hadronic models.}
  \label{tab:result}
\end{table}

\subsection{Analysis methodology}
\label{sec:analysis}

Our analyses include several nuisance parameters that take into account systematic uncertainties in the atmospheric neutrino flux, in the neutrino parameters and in the detector properties. We include nuisance parameters for the normalization of the atmospheric neutrino flux, $N$, for the pion-to-kaon ratio in the atmospheric neutrino flux, $\pi/K$, and for the spectral index of the atmospheric neutrino spectrum, $\Delta \gamma$. Furthermore, we include a nuisance parameter that accounts for uncertainties in the efficiency of the digital optical modules of the detector, DOM$_{\rm eff}$. As for the neutrino parameters, we also take into account the current uncertainties in $\Delta m^2_{31}$ and $\theta_{23}$. In addition, other potentially important systematic errors come from uncertainties in the primary cosmic-ray flux and the hadronic interaction models. Our default choice for most of the results presented below is the combined Honda-Gaisser model and Gaisser-Hillas H3a correction (HG-GH-H3a) for the primary cosmic-ray flux~\cite{Gaisser:2013bla} and the QGSJET-II-4 hadronic model~\cite{Ostapchenko:2010vb}, although we also consider the Zatsepin-Sokolskaya (ZS) flux~\cite{Zatsepin:2006ci} and the SIBYILL2.3 hadronic model~\cite{Riehn:2015oba}. 

The uncertainty on the flux normalization represents an overall normalization of the number of events which affects equally all bins in relative terms, and we allow it to vary freely within a factor of 2 (larger than current uncertainties~\cite{Honda:2006qj, Fedynitch:2012fs}) of the central value. It is important to fit this parameter because it can be significantly different from one, mainly for the ZS primary cosmic-ray flux. The pion-to-kaon ratio affects the relative contribution to the neutrino flux from pion or kaon decays. A larger value $\pi/K$ implies a softer spectrum, as the neutrino flux from kaon decays is harder. We use $\pi/K$ normalized to one and consider a Gaussian prior of 10\%. The uncertainty on the spectral index represents a tilt in the energy spectrum of the atmospheric neutrino flux with a pivot energy near the median of the neutrino energy distribution (so this correction is not very correlated with the normalization), and we apply a Gaussian prior with a 5\% error. Finally, the uncertainty in the optical efficiency affects the determination of the reconstructed energy, so that a larger DOM$_{\rm eff}$ implies a shift to larger energies. For this nuisance parameter we consider a flat prior, which in practice equals to allow it to float freely.

On the other hand, as we discussed above, the high-energy IceCube atmospheric neutrino sample cannot significantly constrain the diagonal NSI parameter $\varepsilon'$, so we constrain this parameter by means of the SK limits, which were obtained using data at lower energies. The SK bound reads~\cite{Mitsuka:2011ty}, 
\beq 
\label{eq:SK}
|\varepsilon'| = | \varepsilon_{\tau \tau} - \varepsilon_{\mu \mu}| < 0.049  ~, \hspace{1cm} \textrm{90\% confidence level (C.L.)} ~,
\eeq
and from Fig.~4 in Ref.~\cite{Mitsuka:2011ty}, we set the $1\sigma$ C.L. prior on $\varepsilon'$ to $\sigma_{\varepsilon'} = 0.040$.

\begin{figure}
	\centering
	\includegraphics[width=\textwidth]{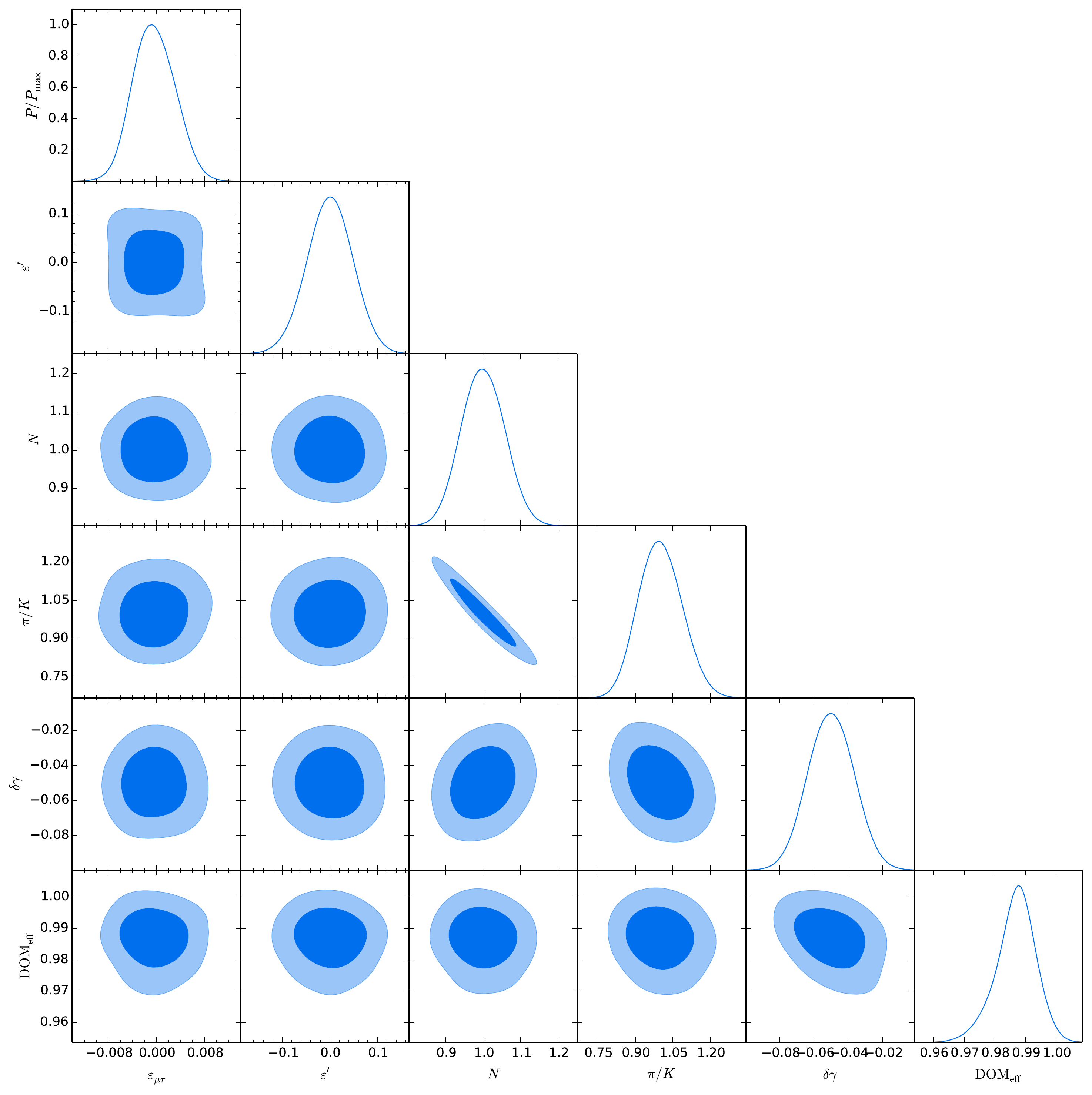}
	\caption{Posterior (68\% and 95\%) probability contours for the NSI parameters $\varepsilon_{\mu \tau}$ and $\varepsilon'$ along with several nuisance parameters, using the one-year through-going muon IceCube data~\cite{TheIceCube:2016oqi}. On the right panels, we also depict the one-dimensional posterior probability distribution of the parameter corresponding to each column. In all the panels we also include the uncertainties on $\Delta m^2_{31}$ and $\theta_{23}$.}
	\label{fig:triangle}
\end{figure}

To quantitatively assess the power of the high-energy atmospheric neutrino one-year IceCube data to constrain NSI in neutrino propagation in matter, we perform a likelihood analysis using all the events in the data sample and characterizing each event by its reconstructed muon energy and zenith angle. The full likelihood is defined as the bin product of the Poisson probability of measuring $N^{\rm data}_i$ for the expected value $N^{\rm th}_i$ times the product of Gaussian probabilities for the pulls of the nuisance parameters. The log-likelihood (up to a constant) is given by
\begin{equation}
\label{eq:L}
\ln \mathcal{L} (\varepsilon_{\mu \tau}, \varepsilon' ; {\boldsymbol \eta}) = \sum_{i\in {\rm bins}}  \left(N^{\rm data}_i \, \ln N^{\rm th}_i (\varepsilon_{\mu \tau}, \varepsilon' ; {\boldsymbol \eta})  - N^{\rm th}_i (\varepsilon_{\mu \tau}, \varepsilon' ; {\boldsymbol \eta})\right) - \frac{\varepsilon'^2}{2 \, \sigma_{\varepsilon'}^2} - \sum_j \frac{(\eta_j-\eta^0_j)^2}{2 \, \sigma_j^2}~,
\end{equation}
where the subindex $i$ refers to a bin in $E^{\rm rec}_\mu$ and $\cos \theta^{\rm rec}_z$, $N^{\rm th}_i (\varepsilon_{\mu \tau}, \varepsilon' ; {\boldsymbol \eta})$ is the expected number of evens for a given value of the NSI ($\varepsilon_{\mu \tau}$ and $\varepsilon'$) and nuisance (${\boldsymbol \eta} \equiv \{N, \pi/K, \Delta \gamma, {\rm DOM}_{\rm eff}, \Delta m^2_{31}, \theta_{23}\}$) parameters in the $i$-th bin, and $N^{\rm data}_i$ is the number of data events in the same $i$-th bin. The index $j$ corresponds to the nuisance parameters with Gaussian prior ($\pi/K$, $\Delta \gamma$, $\Delta m^2_{31}$ and $\theta_{23}$) and $\sigma_j$ is the Gaussian error. To compute the likelihood for a given value of the parameters, we first propagate the neutrino fluxes from the atmosphere to the detector for both neutrinos and antineutrinos, then we weigh the events from the IceCube Monte Carlo with the propagated flux, which is a function of the true neutrino energy $E_\nu$ and the zenith angle $\theta_z$, and we construct two-dimensional histograms as a function of the reconstructed variables: $E^{\rm rec}_\mu$ and $\theta^{\rm rec}_z$ (20 energy and 20 angular bins). With this likelihood, we perform a Bayesian analysis using the MultiNest nested sampling algorithm~\cite{Feroz:2007kg, Feroz:2008xx, Feroz:2013hea} in the NSI and nuisance parameter space. All the parameters, together with their range of variation and the type of prior considered, are summarized in Tab.~\ref{tab:params}.

\subsection{Current bounds}
\label{sec:current}

The results, using our default models for the primary cosmic-ray spectrum and hadronic interactions, are shown in Fig.~\ref{fig:triangle}, where we depict the 68\% and 95\% credible contours (posterior probabilities). Concerning the NSI parameter $\varepsilon_{\mu \tau}$, which is the main goal of this paper, its correlation with the continuous systematic parameters we consider is small. This is somehow expected, as in the ${\cal O}({\rm TeV})$ energy range, the main signature of the presence of matter NSI is via the distortion of the angular distribution of the atmospheric neutrino events and all these systematics mostly affect the atmospheric neutrino energy spectrum, modifying very little its angular distribution. Notice indeed that most of the parameters are not much correlated among themselves, an exception being the pion-to-kaon ratio ($\pi/K$) and the flux normalization ($N$), which show a clear anticorrelation.

\begin{figure}
	\centering
	\includegraphics[width=0.49\textwidth]{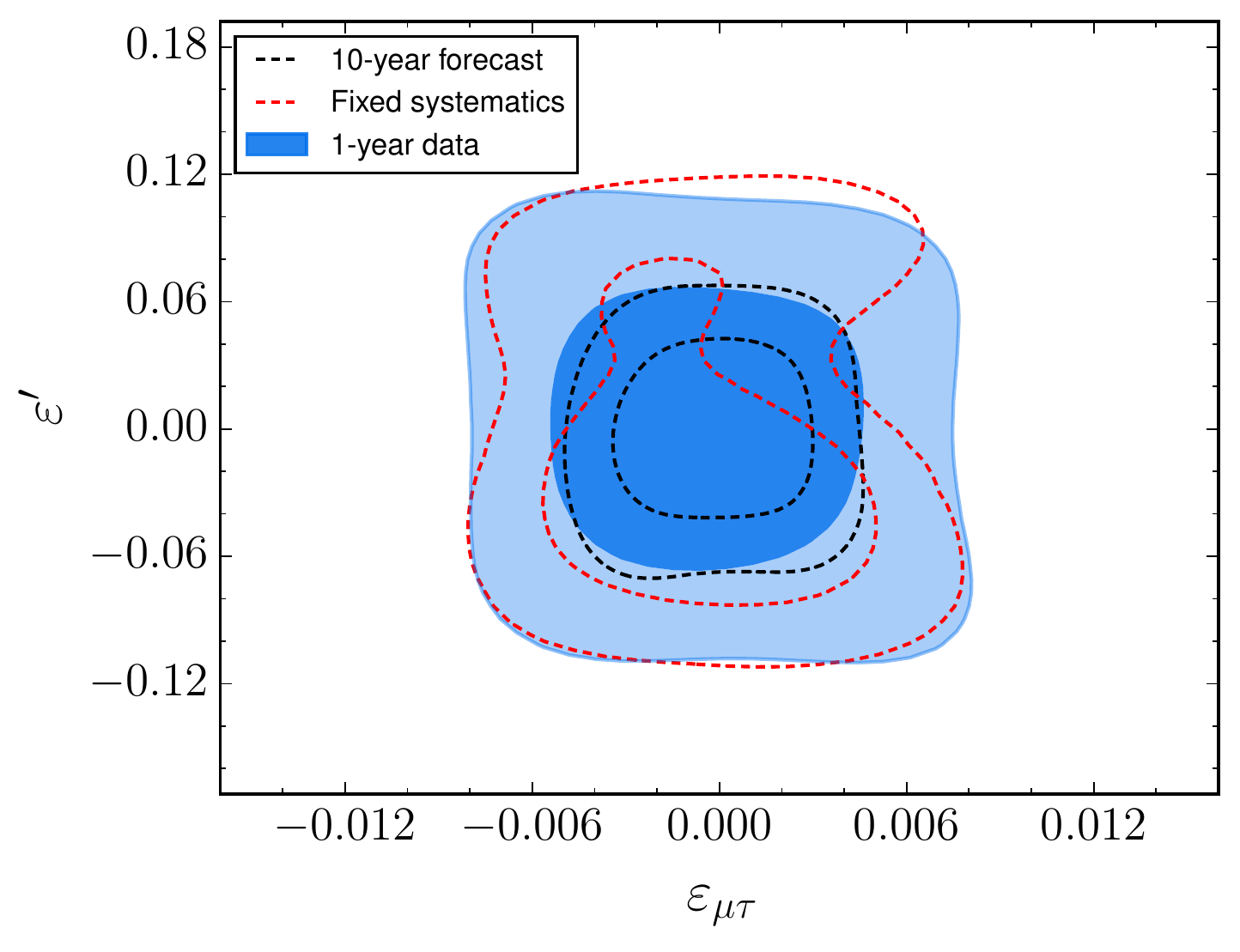}
	\includegraphics[width=0.46\textwidth]{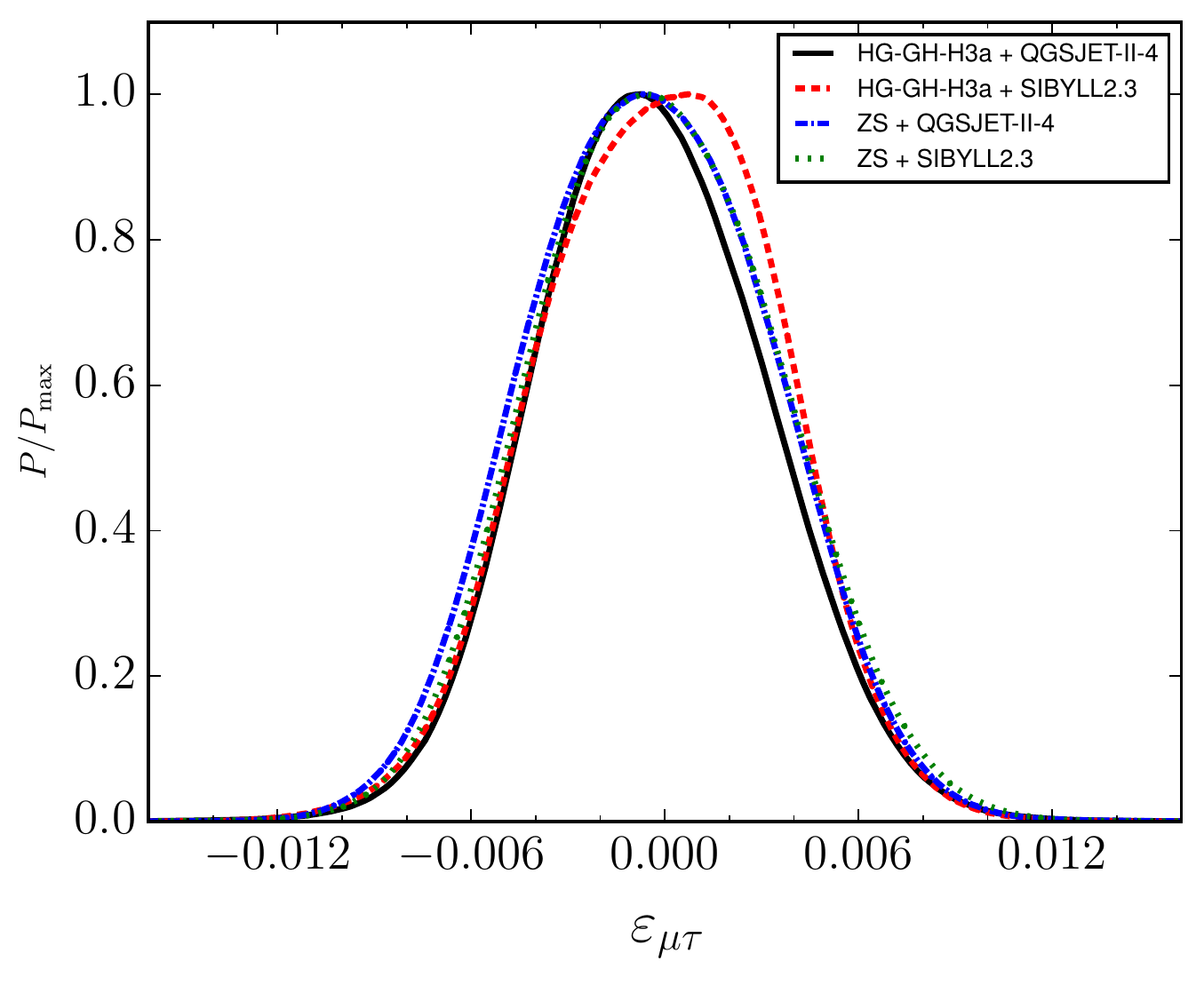}
	\caption{{\it Left panel:} Comparison of the 68\% and 95\% credible contours in the $\varepsilon_{\mu \tau}-\varepsilon'$ plane for our default analysis (filled blue regions) with those obtained when all nuisance parameters are fixed at their default values (red closed curves), see Tab.~\ref{tab:params}. We also show the result expected in the case of no NSI after 10 years of data taking (black closed curves), see Sec.~\ref{sec:future}. {\it Right panel:} Posterior probabilities of $\varepsilon_{\mu \tau}$, after marginalizing with respect to the rest of parameters, for the four combinations of primary cosmic-ray spectrum and hadronic models: our default choice, HG-GH-H3a + QGSJET-II-4 (black solid curve); HG-GH-H3a + SIBYLL2.3 (red dashed curve); ZS + QGSJET-II-4 (blue dot-dashed curve) and ZS + SIBYLL2.3 (green dotted curve).} 
	\label{fig:systematics}
\end{figure}

From this analysis, using the high-energy atmospheric neutrino IceCube data, we obtain the most stringent bound on $\varepsilon_{\mu \tau}$ to date, 
\beq
\label{eq:bound}
- 6.0 \times 10^{-3} < \varepsilon_{\mu \tau} < 5.4 \times 10^{-3} ~, \hspace{1cm} \textrm{90\% credible interval (C.I.)}.
\eeq
The interval is rather symmetric with respect to zero, as the NSI effects depend mainly on $\varepsilon_{\mu \tau}^2$. Our result improves over the SK limit~\cite{Mitsuka:2011ty, Thomas}, 
\beq
\label{eq:SKlimit}
|\varepsilon_{\mu \tau}| < 1.1 \times 10^{-2} \hspace{1cm} \textrm{90\% C.L. (SK)} ~,
\eeq
over the result of a preliminary analysis of three-year DeepCore data~\cite{DayN2016},
\beq
\label{eq:DClimit}
- 6.7 \times 10^{-3} < \varepsilon_{\mu \tau} < 8.1 \times 10^{-3} \hspace{1cm} \textrm{90\% C.L. (DeepCore)} ~,
\eeq
and it is very similar to that obtained in Ref.~\cite{Esmaili:2013fva} using 79-string IceCube configuration and DeepCore data,
\beq
\label{eq:IC79}
- 6.1 \times 10^{-3} < \varepsilon_{\mu \tau} < 5.6 \times 10^{-3} ~, \hspace{1cm} \textrm{90\% C.L. (IC79 + DeepCore)} ~,
\eeq
although note that we have included a number of nuisance parameters not considered in Ref.~\cite{Esmaili:2013fva}.

To further assess the lack of correlation of $\varepsilon_{\mu \tau}$ with the nuisance parameters and the stability of our results with respect to their variation, in the left panel of Fig.~\ref{fig:systematics} we overimpose the contours obtained when fixing all nuisance parameters to their default values (see Tab.~\ref{tab:params}) to those shown in Fig.~\ref{fig:triangle}, where they are varied as described above. It is clear that these systematics affect very little the final bound on $\varepsilon_{\mu \tau}$, which gets modified as
\beq
- 5.1 \times 10^{-3} < \varepsilon_{\mu \tau} < 4.3 \times 10^{-3} ~, \hspace{1cm} \textrm{90\% C.I. (no systematic uncertainties)} ~,
\eeq
for the most optimistic case of not including systematic uncertainties in the analysis. This is a more fair comparison with the results of Ref.~\cite{Esmaili:2013fva}.

We also study the impact of using different primary cosmic-ray spectra and different hadronic interaction models on our results. As discussed above, neutrino NSI in matter may produce a suppression in the high-energy upgoing atmospheric muon data in IceCube, with a characteristic angular dependence (and little energy dependence). Different combinations of primary cosmic-ray spectrum and hadronic models imply slightly different angular distributions for the atmospheric neutrino flux and hence, potentially, they are an important source of systematic uncertainties on the NSI sensitivity reach of neutrino telescopes. In the right panel of Fig.~\ref{fig:systematics} we show the results for different choices of cosmic-ray spectra and hadronic interaction models. We depict the posterior probabilities for $\varepsilon_{\mu \tau}$, marginalized with respect to the rest of the nuisance parameters and $\varepsilon'$, for each of the four possible combinations. Indeed, the allowed range of $\varepsilon_{\mu \tau}$ arising from our default combination of models (HG-GH-H3a + QGSJET-II-4) turns out to be very similar to the resulting ones from all possible combinations, whose bounds on $\varepsilon_{\mu \tau}$ are:

\bea
- 5.5 \times 10^{-3} & < \varepsilon_{\mu \tau} < & 5.1 \times 10^{-3}  ~, \hspace{1cm}  \textrm{90\% C.I. (HG-GH-H3a + SIBYLL2.3)} ~,\\
- 6.0 \times 10^{-3} & < \varepsilon_{\mu \tau} < & 5.1 \times 10^{-3}  ~, \hspace{1cm} \textrm{90\% C.I. (ZS + QGSJET-II-4)} ~, \\
- 6.2 \times 10^{-3} & < \varepsilon_{\mu \tau} < & 5.8 \times 10^{-3}  ~, \hspace{1cm} \textrm{90\% C.I. (ZS + SIBYLL2.3)} ~.
\eea

\begin{figure}
	\centering
	\includegraphics[width=0.78\textwidth]{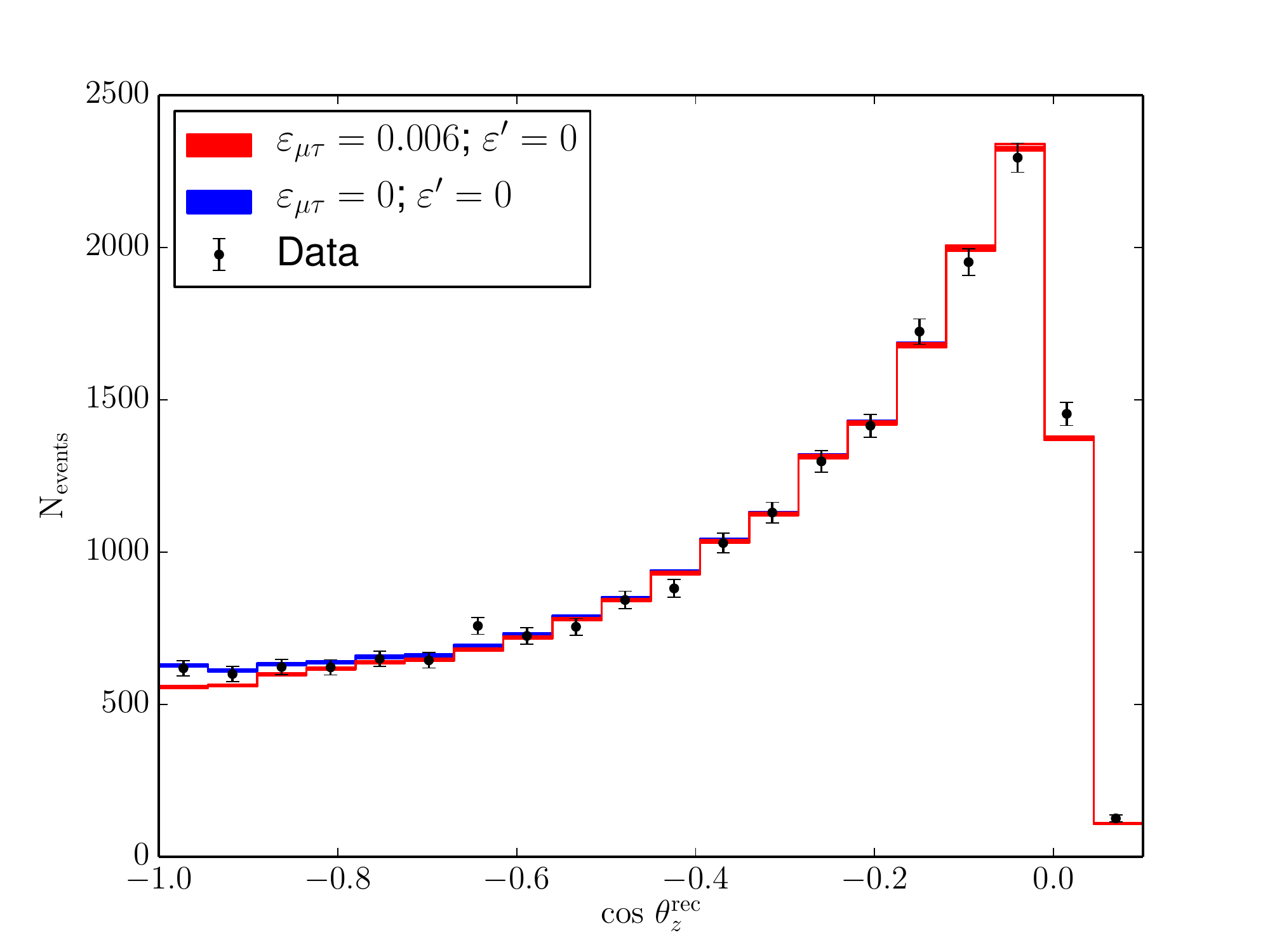}
	\caption{Event spectrum: data points (black dots and error bars), expected results without NSI (red solid histogram) and including NSI with $\varepsilon_{\mu \tau} = 0.006$ and $\varepsilon' = 0$ (blue solid histogram). The uncertainties due to the choice of the primary cosmic-ray spectrum and hadronic models are represented by the width of the histograms.}
	\label{fig:evspectrum}
\end{figure}

Finally, in Fig.~\ref{fig:evspectrum}, we show the event spectrum, integrated in the entire interval in reconstructed muon energy\footnote{As the effect of NSI, at high energies, depends very mildly on the reconstructed muon energy, varying the energy range included in Fig.~\ref{fig:evspectrum} does not change the relevant features.}, as a function of $\cos \theta^{\rm rec}_z$. We show the histogram of the detected through-going atmospheric muon events after one year (black dots), together with their error bars and the expectation for the cases without NSI (red histogram) and when NSI are included with $\varepsilon_{\mu \tau} = 0.006$ and $\varepsilon' = 0$ (blue histogram). We indicate the uncertainty due to the choice of the primary cosmic-ray spectrum and hadronic models by the width of the histograms, although the variation is very small. In all cases we consider the best fit values for the parameters. As discussed in previous sections, we see that the presence of NSI implies a suppression of the atmospheric neutrino flux, and hence the observed through-going muon spectra, for neutrinos crossing the core of the Earth, i.e., for $\cos \theta_z \gtrsim -0.8$. 

All our results are summarized in Tab.~\ref{tab:result}.

\subsection{Forecast analyses}
\label{sec:future}

\begin{figure}
	\centering
	\includegraphics[width=\textwidth]{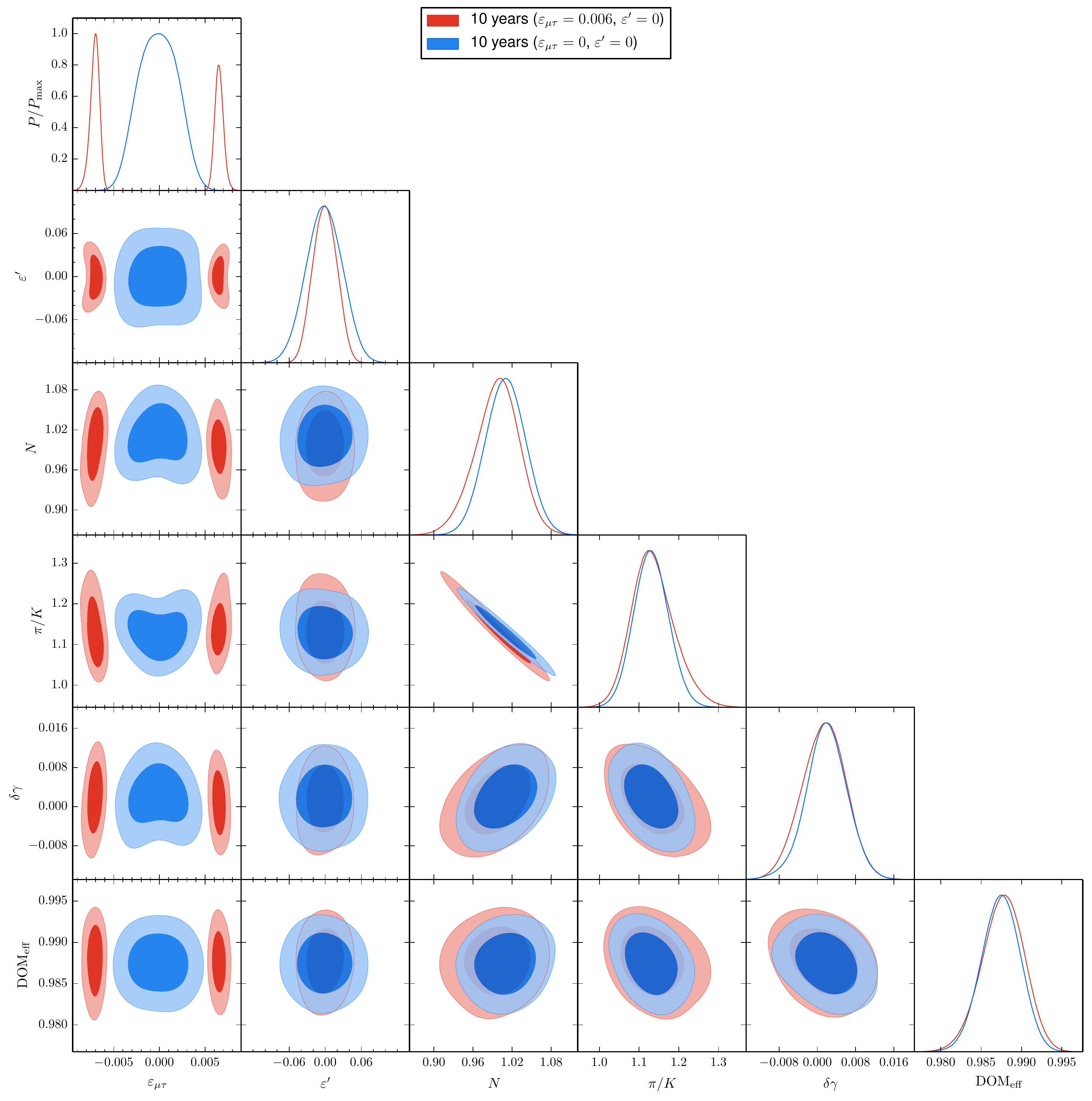}
	\caption{Posterior (68\% and 95\%) probability contours for two 10-year forecasts of high-energy atmospheric neutrino data in IceCube. We show the results assuming the data corresponds to the case without NSI (blue contours) and when the data includes NSI with $\varepsilon_{\mu \tau}=0.006$ and $\varepsilon' = 0$ (red contours). On the right panels, we also show the one-dimensional probability distribution of the parameter corresponding to each column. The atmospheric neutrino parameters $\Delta m^2_{31}$ and $\theta_{23}$ are fixed to their current best fit values.}
	\label{fig:triangle10}
\end{figure}

Finally, in order to assess the capabilities of the IceCube detector when using the high-energy atmospheric neutrino data to constrain matter NSI, we perform two forecast analyses for 10 years of simulated data. Therefore, we simulate 10 years of data and use the same priors on all parameters as described above, except from $\Delta m^2_{31}$ and $\theta_{23}$ which we fix to their best fit values and from $\varepsilon'$ which we take $\sigma_{\varepsilon'} = 0.03$, and our default combination of primary cosmic-ray and hadronic interaction models (HG-GH-H3a + QGSJET-II-4). The results are shown in Fig.~\ref{fig:triangle10} and in the left panel of Fig.~\ref{fig:systematics}. 

On one hand, we simulate data assuming the case without NSI to be the true case realized in Nature (blue contours). As expected from current bounds (see Fig.~\ref{fig:triangle}), the systematics described by the nuisance parameters we consider do not play a significant role for future analyses, although some small correlations start to show up more clearly, which partially limit the expected reach. After 10 years of collecting data, we would expect $\varepsilon_{\mu \tau}$ to be constrained in the interval
\beq
\label{eq:forecast}
- 3.3 \times 10^{-3} < \varepsilon_{\mu \tau} < 3.0 \times 10^{-3} \hspace{1cm} \textrm{90\% C.I. (10-year forecast analysis).}
\eeq
Note that, with a factor of 10 more statistics, the improvement in the limits on NSI will be of about a factor of two. This can also be seen in the left panel of Fig.~\ref{fig:systematics}, where we show the 68\% and 95\% credible contours in the $\varepsilon_{\mu \tau}-\varepsilon'$ plane (black closed curves).

On the other hand, it is also interesting to consider the discovery potential in case of the presence of matter NSI. In order to do this, we simulate 10 years of data including NSI assuming that Nature has chosen $\varepsilon_{\mu \tau} = 0.006$  (which represents a value allowed by current data with about 90\% probability) and $\varepsilon' = 0$. As it is clear from Fig.~\ref{fig:triangle10}, for this large value of $\varepsilon_{\mu \tau}$ (red contours), future IceCube measurements would be able to detect the presence of matter NSI at a high significance, although the quadratic $\varepsilon_{\mu \tau}$-dependence of the effects would render impossible to determine the sign of $\varepsilon_{\mu \tau}$.

\section{Summary and conclusions}
\label{sec:conclusions}

The IceCube neutrino telescope, along with its low-energy extension DeepCore, is currently the leading experiment to detect high-energy neutrinos. After a few years of operation, statistics have been accumulated and a number of studies of atmospheric neutrinos have been performed. Atmospheric neutrino flux measurements have been carried out in a wide range of energies~\cite{Abbasi:2010ie, Aartsen:2012uu, Aartsen:2014qna, Aartsen:2014muf, Aartsen:2015xup, Aartsen:2015nss}, low-energy atmospheric neutrino data have been considered to constrain neutrino oscillation parameters to levels comparable to other neutrino oscillation experiments~\cite{Aartsen:2013jza, Aartsen:2014yll}, low-energy data have also been used to set constraints on matter NSI~\cite{DayN2016}, and light sterile neutrinos, claimed as possible explanations of short baseline neutrino anomalies~\cite{Athanassopoulos:1996jb, Aguilar:2001ty, AguilarArevalo:2007it, AguilarArevalo:2010wv, Aguilar-Arevalo:2013pmq, Acero:2007su, Giunti:2010zu, Mention:2011rk}, have been searched for with atmospheric neutrinos with energies up to ${\cal O}(10~{\rm TeV})$~\cite{TheIceCube:2016oqi}.

Although there are still unknowns within the standard picture of neutrino oscillations, new interactions in the neutrino sector, driven by dimension six (or higher) operators, could also give rise to sub-leading effects in neutrino production, propagation and detection. In this work we have considered the high-energy atmospheric neutrino data (previously used to search for sterile neutrinos in Ref.~\cite{TheIceCube:2016oqi}), i.e., high-energy through-going muon events, to evaluate the impact of matter NSI in the neutrino propagation through the Earth. Our analysis represents the most stringent limit on the $\mu\tau$-sector off-diagonal parameter $\varepsilon_{\mu \tau}$ to date and it is complementary to other studies in the literature, which focus on atmospheric neutrino events at lower energies~\cite{Fornengo:2001pm, GonzalezGarcia:2004wg, Friedland:2004ah, Friedland:2005vy, GonzalezGarcia:2011my, Mitsuka:2011ty, Esmaili:2013fva, Mocioiu:2014gua, Fukasawa:2015jaa, Thomas, Day:2016shw, DayN2016} and to the results obtained using the 79-string IceCube configuration~\cite{Esmaili:2013fva}. 

We have first reviewed the formalism of matter NSI, which is relevant for high-energy atmospheric neutrinos (Sec.~\ref{sec:forma}). Although we have computed neutrino propagation in a full three-neutrino framework taking into account Earth attenuation and degradation in energy due to NC interactions, at the energies we consider, the description in terms of a two-neutrino system represents a very good approximation, as the (mostly) $\nu_e$ state decouples. Therefore, we have described the main features within this approximation as two-neutrino oscillations in the $\mu\tau$-sector. We have illustrated the effect of flavor transitions, with and without NSI, and attenuation by depicting ratios of propagated to unpropagated neutrino and antineutrino fluxes (left panels of Figs.~\ref{fig:fratio_1D} and~\ref{fig:fratio_2D}) and have also isolated the effect of NSI by considering ratios of propagated fluxes with and without NSI (right panels of Figs.~\ref{fig:fratio_1D} and~\ref{fig:fratio_2D}). 

Then, we have briefly described the high-energy upgoing muon sample used to perform our analysis (Sec.~\ref{sec:data}) and, to understand the features previously discussed at the level of fluxes, we have also studied their impact on the current observables, which cannot distinguish neutrinos from antineutrinos. In order to do so, we have simulated mock data and have shown the difference of expected number of events with and without NSI (left panel of Fig.~\ref{fig:DeltaEv}) and the statistical pulls of NSI effects (right panel of Fig.~\ref{fig:DeltaEv}) as a function of observables: reconstructed muon energy and zenith angle. As the energy dependence of the NSI effects for neutrinos and antineutrinos either tend to cancel out (at low energies in our sample) when both contributions are summed up, or are the same (at higher energies), the dominant distortion of the total event spectrum occurs in the angular distribution (Fig.~\ref{fig:DeltaEvfrac}). 

In the likelihood we define to perform our statistical analysis (Sec.~\ref{sec:analysis}), we have also included several nuisance parameters to describe systematic uncertainties on the atmospheric neutrino flux, on the neutrino parameters and on the detector properties, and have used a prior on the value of the diagonal NSI parameter $\varepsilon'$ from SK measurements (Tab.~\ref{tab:params}). Our results (Sec.~\ref{sec:current}) turn out not to be very correlated with any of the continuous parameters we consider (Fig.~\ref{fig:triangle}) and we obtain a limit on the off-diagonal term $\varepsilon_{\mu \tau}$ (Tab.~\ref{tab:result}). The bound, for the combination of primary cosmic-ray and hadronic models HG-GH-H3a + QGSJET-II-4, is 
\beq
- 6.0 \times 10^{-3} < \varepsilon_{\mu \tau} < 5.4 \times 10^{-3} \hspace{1cm} \textrm{90\% C.I.} ~, 
\eeq
which represents the most stringent limit on this parameter to date.

This limit is very stable with respect to all the continuous nuisance parameters we consider, which can be safely fixed to their default values without affecting significantly the bound (left panel of Fig.~\ref{fig:systematics}). On the other hand, we find the main source of systematic uncertainties to lie in the choice of the combination of primary cosmic-ray and hadronic interaction models (right panel of Fig.~\ref{fig:systematics}). This is explained by the angular dependence of the event spectrum, which is slightly different for each of these combinations. We have also shown this uncertainty by depicting the event spectrum with and without NSI (Fig.~\ref{fig:evspectrum}) as a function of the zenith angle, accounting for the range of the four combinations we consider. Finally, we have also performed a forecast with simulated data for 10 years in IceCube (Sec.~\ref{sec:future}) and noted that, although limits will not improve dramatically (within a factor of two), in case of the existence of large NSI, consistent with current limits, IceCube high-energy atmospheric neutrino data can establish its presence at high confidence (Fig.~\ref{fig:triangle10}). 

Unveiling the values of the parameters of the neutrino sector with high precision requires experimental setups that allow us to test different ranges of energies and baselines. In particular, it requires high-precision measurements, sensitive to non-canonical, sub-leading effects, as those caused by the potential existence of NSI. Neutrino telescopes as IceCube are sensitive to a wide range of energies and baselines and provide an excellent tool to explore possible new neutrino interactions beyond the standard neutrino oscillation paradigm by means of high-energy atmospheric neutrinos. Higher statistics will allow us to further test this scenario and in this regard, a future high-energy extension of the IceCube detector~\cite{Aartsen:2014njl} and the planned KM3NeT telescope~\cite{Adrian-Martinez:2016fdl} will have a crucial role.

\section*{Acknowledgments}
The authors are supported by the Generalitat Valenciana under Grants PROMETEOII/2014/049 and PROMETEO II/2014/050, by the Spanish Grants FPA2014--57816-P, FPA2014-54459-P and SEV-2014-0398 of the MINECO, and by the European Union's Horizon 2020 research and innovation program under the Marie Sk\l odowska-Curie grant agreements No. 690575 and 674896. SPR is supported by a Ram\'on y Cajal contract, and also partially by the Portuguese FCT through the CFTP-FCT Unit 777 (PEst-OE/FIS/UI0777/2013).

\bibliographystyle{apsrev4-1}
\bibliography{nsi_ice}

\end{document}